\documentclass[10pt]{iopart}

\usepackage{braket}
\usepackage{graphicx}
\usepackage{amssymb}
\usepackage[colorlinks=true,linkcolor=blue,urlcolor=blue,citecolor=blue]{hyperref}

\begin{document}

\title[Fermion Sign Problem in PIMC Simulations: Grand-canonical ensemble]{Fermion Sign Problem in Path Integral Monte Carlo Simulations: Grand-canonical ensemble}

\author{Tobias Dornheim}

\address{Center for Advanced Systems Understanding (CASUS), D-02826 G\"orlitz, Germany\\
Helmholtz-Zentrum Dresden-Rossendorf (HZDR), D-01328 Dresden, Germany}
\ead{t.dornheim@hzdr.de}
\vspace{10pt}
\begin{indented}
\item[]April 2021
\end{indented}

\begin{abstract}
We present a practical analysis of the fermion sign problem in fermionic path integral Monte Carlo (PIMC) simulations in the grand-canonical ensemble (GCE). As a representative model system, we consider electrons in a $2D$ harmonic trap. We find that the sign problem in the GCE is even more severe than in the canonical ensemble at the same conditions, which, in general, makes the latter the preferred option. Despite these difficulties, we show that fermionic PIMC simulations in the GCE are still feasible in many cases, which potentially gives access to important quantities like the compressiblity or the Matsubara Greens function. This has important implications for contemporary fields of research such as warm dense matter, ultracold atoms, and electrons in quantum dots.
\end{abstract}

%
%
%
%
%

\section{Introduction}

Having originally been introduced for the description of $^4$He in the 1960s~\cite{Fosdick_PR_1966,Jordan_PR_1968}, the path integral Monte Carlo (PIMC) approach~\cite{Berne_JCP_1982,Takahashi_Imada_PIMC_1984,Pollock_Ceperley_PRB_1984,cep} constitutes one of the most successful methods in statistical physics and related disciplines. More specifically, the PIMC method in principle allows to obtain quasi-exact results for quantum many-body systems at finite temperature without any empirical external input. This has allowed for unprecedented insights into important physical phenomena such as superfluidity~\cite{cep,ultracold2,Dornheim_PRB_2015,Kwon_PRL_2002}, Bose-Einstein-condensation~\cite{Saito,PhysRevLett.56.351}, and even collective excitations~\cite{Saccani_Supersolid_PRL_2012,Filinov_PRA_2012,dornheim_dynamic,PhysRevB.98.134509}.

A particular strength of the PIMC method is its capability to exactly resolve the nontrivial interplay between coupling effects, thermal excitations, and quantum statistics, as it occurs in fields like warm dense matter~\cite{review,new_POP,wdm_book,fortov_review}, electrons in quantum dots~\cite{Egger_PRL_1999,RevModPhys.74.1283}, and ultracold atoms~\cite{Filinov_PRA_2012,Dornheim_PRA_2020}. For bosons (or hypothetical distinguishable particles, often called \emph{Boltzmannons}), modern sampling techniques~\cite{boninsegni1,boninsegni2} allow for PIMC simulations of up to $N\sim10^4$ particles. Yet, the description of e.g. warm dense matter requires the simulation of electrons, which are fermions. The antisymmetry of the fermionic density matrix under the exchange of particles leads to the notorious fermion sign problem (FSP)~\cite{Loh_sign_problem_PRB_1990,dornheim_sign_problem}, which is one of the most fundamental and challenging obstacles in theoretical physics and quantum chemistry.
More specifically, the FSP leads to an exponential increase in compute time towards low temperature and with increasing system size, and it has been formally shown to be $NP$-hard for some cases~\cite{troyer}.

The high current need for an accurate description of quantum degenerate nonideal fermions at finite temperature has sparked a surge of activity in the field of quantum Monte Carlo simulations in thermal equilibrium~\cite{Brown_PRL_2013,Blunt_PRB_2014,Dornheim_NJP_2015,Blunt_PRL_2015,Malone_PRL_2016,dornheim_prl,groth_prl,Yilmaz_JCP_2020,Lee_JCP_2021,Dornheim_PRL_2020,dornheim_ML,Rubenstein_auxiliary_finite_T}. Yet, a more practical perspective on the manifestation of the sign problem itself has long remained missing in the respective literature. Recently, Dornheim~\cite{dornheim_sign_problem} has presented an extensive yet accessible investigation of the FSP of fermionic PIMC simulations in the canonical ensemble (CE), i.e., for a fixed number density $n=N/V$.
In the present work, we extend this analysis for the case of a grand-canonical ensemble (GCE), where both $N$ and $n$ can fluctuate in the simulation. On the one hand, the GCE allows for the computation of interesting properties such as the compressibility and the Matsubara Green function~\cite{Filinov_PRA_2012,boninsegni1}, which are not directly accessible in the CE. On the other hand, we show that the GCE leads to a comparably more severe FSP, and to a potential sampling problem in the PIMC simulation due to the divergence of particle number distributions of bosons and fermions. Nevertheless, we find that fermionic PIMC simulations in the GCE are still feasible in some cases, which has important implications for the research fields mentioned earlier.

The paper is organized as follows: in Sec.~\ref{sec:theory}, we introduce the relevant theoretical background, including the PIMC method (\ref{sec:PIMC}), the fermion sign problem (\ref{sec:FSP}), the grand-canonical ensemble (\ref{sec:GCE}), and the model system employed throughout this work (\ref{sec:Hamiltonian}).
Sec.~\ref{sec:results} is devoted to the presentation of our new PIMC results in the GCE, where we investigate the dependence of the FSP on the temperature (\ref{sec:temperature}), the chemical potential (\ref{sec:mu}), and the coupling strength (\ref{sec:coupling}). In addition, we discuss the distributions of Monte Carlo expectation values (\ref{sec:distribution}), which strongly deviate from the usual Gaussian in the case of fermions. The paper is concluded by a brief summary and discussion in Sec.~\ref{sec:summary}.

\section{Theory\label{sec:theory}}

\subsection{Path integral Monte Carlo\label{sec:PIMC}}

The basic idea of the path integral Monte Carlo method~\cite{cep,Takahashi_Imada_PIMC_1984,Berne_JCP_1982} is to express the expectation value of an, in principle, arbitrary observable $\hat O$ in coordinate space,
\begin{eqnarray}\label{eq:Trace}
\braket{\hat O} = \frac{1}{Z} \textnormal{Tr}\hat\rho\hat O = \frac{1}{Z} \int \textnormal{d}\mathbf{R} \bra{\mathbf{R}} \hat\rho \hat O \ket{\mathbf{R}}\ ,
\end{eqnarray}
where $\hat\rho$ ($Z$) is the either canonical or grandcanonical density operator (partition function), and $\mathbf{R}=(\mathbf{r}_1,\dots,\mathbf{r}_N)^T$ contains the coordinates of all particles. We note that the particle number $N$ is allowed to vary in the case of the GCE, see Sec.~\ref{sec:GCE} below for more details. While the direct evaluation of Eq.~(\ref{eq:Trace}) is, in general, not feasible, this equation can be re-cast into a high-dimensional integral over a function that can be readily evaluated---the eponymous \emph{path integral}---which is then evaluated stochastically to avoid the well-known curse of dimensionality of standard quadrature methods.

For identical particles such as bosons or fermions, Eq.~(\ref{eq:Trace}) needs to be extended by the sum over all possible permutations of particle coordinates, which gives in the canonical ensemble
\begin{eqnarray}\label{eq:Trace2}
\braket{\hat O} = \frac{1}{Z}\frac{1}{N!} \sum_{\sigma\in S_N} \textnormal{sgn}\left(\sigma\right) \int \textnormal{d}\mathbf{R} \bra{\mathbf{R}} \hat\rho \hat O \ket{\hat\pi_{\sigma}\mathbf{R}}\ .
\end{eqnarray}
Here the sum is taken over all elements $\sigma$ of the permutation group $S_N$ with $\hat\pi_{\sigma}$ being the corresponding permutation operator. For bosons, it holds $\textnormal{sgn}(\sigma)=1$ and all terms are positive definitive. Hence, The expectation value from Eq.~(\ref{eq:Trace2}) can always be rewritten as 
\begin{eqnarray}\label{eq:weights}
\braket{\hat O} = \frac{1}{Z} \int \textnormal{d}\mathbf{X}\ W(\mathbf{X}) O(\mathbf{X})\ ,
\end{eqnarray}
where $\mathbf{X}$ is a so-called \emph{configuration}, $W(\mathbf{X})$ denotes the corresponding \emph{configuration weight}, and $O(\mathbf{X})$ is the associated estimator of the observable $\hat O$.
Furthermore, the function $P(\mathbf{X})=W(\mathbf{X})/Z$ constitutes a real probability density, as
\begin{eqnarray}
\int \textnormal{d}\mathbf{X}\ P(\mathbf{X}) = 1\ .
\end{eqnarray}
The basic workflow of the PIMC method is to use the Metropolis algorithm~\cite{metropolis} to sample a Markov chain of $N_\textnormal{MC}$ random configurations $\{\mathbf{X}_i\}$ that are distributed according to $P(\mathbf{X})$.
The Monte Carlo estimation of the expectation value of $\hat O$ is then straightforwardly computed as
\begin{eqnarray}
\braket{\hat O}_{N_\textnormal{MC}} = \frac{1}{N_\textnormal{MC}} \sum_{i=1}^{N_\textnormal{MC}} O(\mathbf{X}_i)\ ,
\end{eqnarray}
which becomes exact in the limit of a large number of samples $N_\textnormal{MC}$,
\begin{eqnarray}
\lim_{N_\textnormal{MC}\to\infty} \braket{\hat O}_{N_\textnormal{MC}} = \braket{\hat O}\ .
\end{eqnarray}
In addition, it is known from the central limiting theorem that $\braket{\hat O}_{N_\textnormal{MC}}$ is a random number that is normally distributed around the exact expectation value $\braket{\hat O}$, and the corresponding variance is given by 
\begin{eqnarray}
\Delta O = \frac{\sigma_O}{\sqrt{N_\textnormal{MC}}}\ .
\end{eqnarray}
Here $\sigma_O$ is the intrinsic variance of the estimator $O(\mathbf{X})$, and can be estimated from the $N_\textnormal{MC}$ Monte Carlo samples as
\begin{eqnarray}\label{eq:var}
\sigma_O = \left(
\frac{1}{N_\textnormal{MC}}\sum_{i=1}^{N_\textnormal{MC}}\left(O(\mathbf{X}_i)- \braket{\hat O}_{N_\textnormal{MC}}  \right)^2
\right)^{1/2}\ .
\end{eqnarray}
In a nutshell, the statistical uncertainty $\Delta O$ of the Monte Carlo estimation of an arbitrary observable $\hat O$ can be decreased by increasing the number of Monte Carlo samples as $\Delta O\sim1/\sqrt{N_\textnormal{MC}}$. Therefore, such Monte Carlo methods are often denoted as \emph{quasi-exact}.

In the next section, we turn our attention to the case of fermions, giving rise to the notorious fermion sign problem.

\subsection{Fermion sign problem\label{sec:FSP}}
For fermions, the sign function in Eq.~(\ref{eq:Trace2}) can be both positive and negative, depending on the number of pair permutations $p_\sigma$ of each permutation element $\sigma$,
\begin{eqnarray}
\textnormal{sgn}(\sigma) = (-1)^{p_\sigma}\ .
\end{eqnarray}
Consequently, the weights in Eq.~(\ref{eq:weights}), too, can be negative, and $P(\mathbf{X})=W(\mathbf{X})/Z$ is not a real probability.

To circumvent this issue, one can use the Metropolis algorithm to sample configurations $\mathbf{X}$ from the modified distribution
\begin{eqnarray}
P'(\mathbf{X}) = \frac{W'(\mathbf{X})}{Z'} = \frac{|W(\mathbf{X})|}{Z'}\ ,
\end{eqnarray}
with $Z'$ being the modified normalization
\begin{eqnarray}
Z' = \int\textnormal{d}\mathbf{X}\ W'(\mathbf{X})\ .
\end{eqnarray}
In the case of the standard PIMC method in coordinate space, i.e., without antisymmetrized imaginary-time propagators~\cite{Dornheim_CPP_2019}, this modified configuration space directly corresponds to a bosonic simulation discussed above.
The exact fermionic expectation value of the observable $\hat O$ is then computed as
\begin{eqnarray}\label{eq:ratio}
\braket{\hat O} = \frac{\braket{\hat O\hat S}'}{\braket{\hat S}'}\ ,
\end{eqnarray}
where $S(\mathbf{X})=W(\mathbf{X})/|W(\mathbf{X})|$ measures the sign of the fermionic weight function of a configuration $\mathbf{X}$.
In other words, to calculate an expectation value for a Fermi-system, we simulate a corresponding Bose-system at the same conditions and subsequently extract the desired information from Eq.~(\ref{eq:ratio}). The denominator is commonly known simply as the \emph{average sign} $S$ and constitutes a straightforward measure for the amount of cancellation of positive and negative terms due to the fermionic antisymmetry under particle exchange. In fact, it is well known~\cite{dornheim_POP} that $S$ exponentially decreases both upon increasing the system size $N$ or the inverse temperature $\beta=1/k_\textnormal{B}T$,
\begin{eqnarray}\label{eq:exponential}
S = e^{-\beta N \Delta f}\ ,
\end{eqnarray}
where $\Delta f = f-f'$ is the difference in the free energy per particle of the actual and the modified system.
This is highly problematic, as the statistical uncertainty of the ratio of expectation values in Eq.~(\ref{eq:ratio}) is inversely proportional to $S$, i.e.,
\begin{eqnarray}\label{eq:MC_error}
\frac{\Delta A}{A}\sim \frac{1}{S\sqrt{N_\textnormal{MC}}}\ .
\end{eqnarray}
In other words, there appears an \emph{exponential wall} in fermionic PIMC simulations with respect to $N$ and $\beta$, which can only be compensated by increasing the amount of compute time as $\sim 1/\sqrt{N_\textnormal{MC}}$, which inevitably becomes infeasible. This issue is known as the \emph{fermion sign problem} throughout the literature~\cite{dornheim_sign_problem,Loh_sign_problem_PRB_1990,troyer,dornheim_permutation_cycles}, and has been revealed as $NP$-hard for some cases by Troyer and Wiese~\cite{troyer}.
For example, when the temperature $T$ goes to zero, both the enumerator and the denominator of Eq.~(\ref{eq:ratio}) vanish simultaneously and $\Delta A/A$ diverges~\cite{krauth2006statistical}.

\subsection{Grand canonical ensemble\label{sec:GCE}}

Let us next focus on the task at hand, i.e., fermionic PIMC simulations in the GCE.
The grand canonical partition function is given by
\begin{eqnarray}
Z_\textnormal{GC} = \sum_{N=1}^\infty e^{-\beta\mu N} Z_N\ ,
\end{eqnarray}
where $\mu$ is the chemical potential and $Z_N$ is the canonical partition function for $N$ particles; see e.g. Ref.~\cite{dornheim_permutation_cycles} for an explicit expression for $Z_N$ in the context of the PIMC method.

Obviously, $N$ is not constant and varies throughout the PIMC simulation.
For bosons, the particle number distribution is simply given by
\begin{eqnarray}\label{eq:PN}
P(N) &=& \braket{\delta_{{\hat N},N}}\\ \nonumber
&=& \frac{N_N}{N_\textnormal{MC}}\ ,
\end{eqnarray}
where $N_N$ is the number of configurations with particle number $N$. Furthermore, from now on we always assume $\braket{\dots}$ to denote a grand-canonical expectation value unless stated otherwise.

For a Monte Carlo simulation with a sign problem, Eq.~(\ref{eq:PN}) is transformed to 
\begin{eqnarray}
P(N) = \frac{\braket{\delta_{{\hat N},N}\hat S}'}{\braket{\hat S}'}\ ,
\end{eqnarray}
which can be straightforwardly rewritten as
\begin{eqnarray}\label{eq:PN_fermi}
P(N) = \frac{N_N}{N_\textnormal{MC}} \frac{\braket{\hat S}'_{N}}{\braket{\hat S}'}\ .
\end{eqnarray}
Evidently, the main impact of the sign problem on the particle number distribution is a re-weighting of $P(N)$ by the corresponding canonical sign $\braket{\hat S}'_{N}$, with the total grand canonical sign $\braket{\hat S}'$ being the normalization. In addition, we note that the exponential decrease of the canonical sign with $N$, see Eq.~(\ref{eq:exponential}), means that the actual fermionic distribution will be shifted towards smaller $N$ compared to the effective Bose-system which we actually simulate.

Considering the total grand canonical sign itself, we find
\begin{eqnarray}\label{eq:GC_sign}
\braket{\hat S}' = \sum_{N=1}^\infty \left( 
\frac{N_N}{N_\textnormal{MC}} \braket{\hat S}'_{N}
\right)\ .
\end{eqnarray}
In other words, $\braket{\hat S}'$ is given as a sum over the respective canonical signs, weighted by the actual bosonic particle number distribution, Eq.~(\ref{eq:PN}). Thus, Eq.~(\ref{eq:GC_sign}) immediately indicates that a fermionic PIMC simulation in the GCE with $\braket{\hat N}=N$ is afflicted with a substantially more severe sign problem compared to a corresponding simulation in the CE with $N$ being fixed. The first reason for this behaviour is the exponential decrease of $\braket{\hat S}'_{N}$ with $N$, which indicates that configurations with $N>\braket{\hat N}$ lead to a disproportional decrease of the grand canonical sign. The second reason is the fact that the contributions to Eq.~(\ref{eq:GC_sign}) of individual $N$ are weighted not by their actual fermionic distribution given in Eq.~(\ref{eq:PN_fermi}), but by the bosonic distribution occurring in the PIMC simulation itself. Thus, large particle numbers $N>\braket{\hat N}$ with a practically negligible fermionic weight $P(N)$ can potentially strongly decrease the average sign, and therefore strongly increase the statistical uncertainty, cf.~Eq.~(\ref{eq:MC_error}).

\subsection{Model system\label{sec:Hamiltonian}}

We consider $N$ spin-polarized electrons in a strictly $2D$ harmonic oscillator potential potential,
\begin{eqnarray}\label{eq:Hamiltonian_trap}
\hat H = - \frac{1}{2} \sum_{k=1}^N \nabla_k^2 + \frac{1}{2} \sum_{k=1}^N \mathbf{\hat r}_k^2 + \sum_{k>l}^N \frac{ \lambda }{ |\mathbf{\hat r}_l - \mathbf{\hat r}_k| } \quad ,
\end{eqnarray}
which is often used as a convenient model system for the development and benchmark of quantum Monte Carlo methods and related approaches~\cite{Dornheim_NJP_2015,https://doi.org/10.1002/ctpp.201100012,doi:10.1063/5.0030760,doi:10.1063/5.0008720}.

We assume oscillator units, corresponding to the characteristic length $l_0=\sqrt{\hbar/m\Omega}$ (with $\Omega$ being the trap frequency) and energy scale $E_0=\hbar\Omega$. The first term corresponds to the kinetic contribution $\hat K$ and the last two terms to the external potential and the Coulomb interaction, $\hat V_\textnormal{ext}$ and $\hat W$, respectively. For completeness, we mention that the Hamiltonian Eq.~(\ref{eq:Hamiltonian_trap}) is often used as a simple model for electrons in a quantum dot~\cite{RevModPhys.74.1283}.

\section{Results\label{sec:results}}

All PIMC results in this work have been obtained using an implementation of the worm algorithm by Boninsegni \textit{et al.}~\cite{boninsegni1,boninsegni2}, which automatically operates in the GCE. Furthermore, we use a primitive factorization of the density matrix and the convergence with the number of imaginary-time steps has been carefully checked; see the appendix of Ref.~\cite{dornheim_sign_problem} for a corresponding analysis for a similar system.

\subsection{Temperature dependence\label{sec:temperature}}

\begin{figure}\centering
\includegraphics[width=0.5\textwidth]{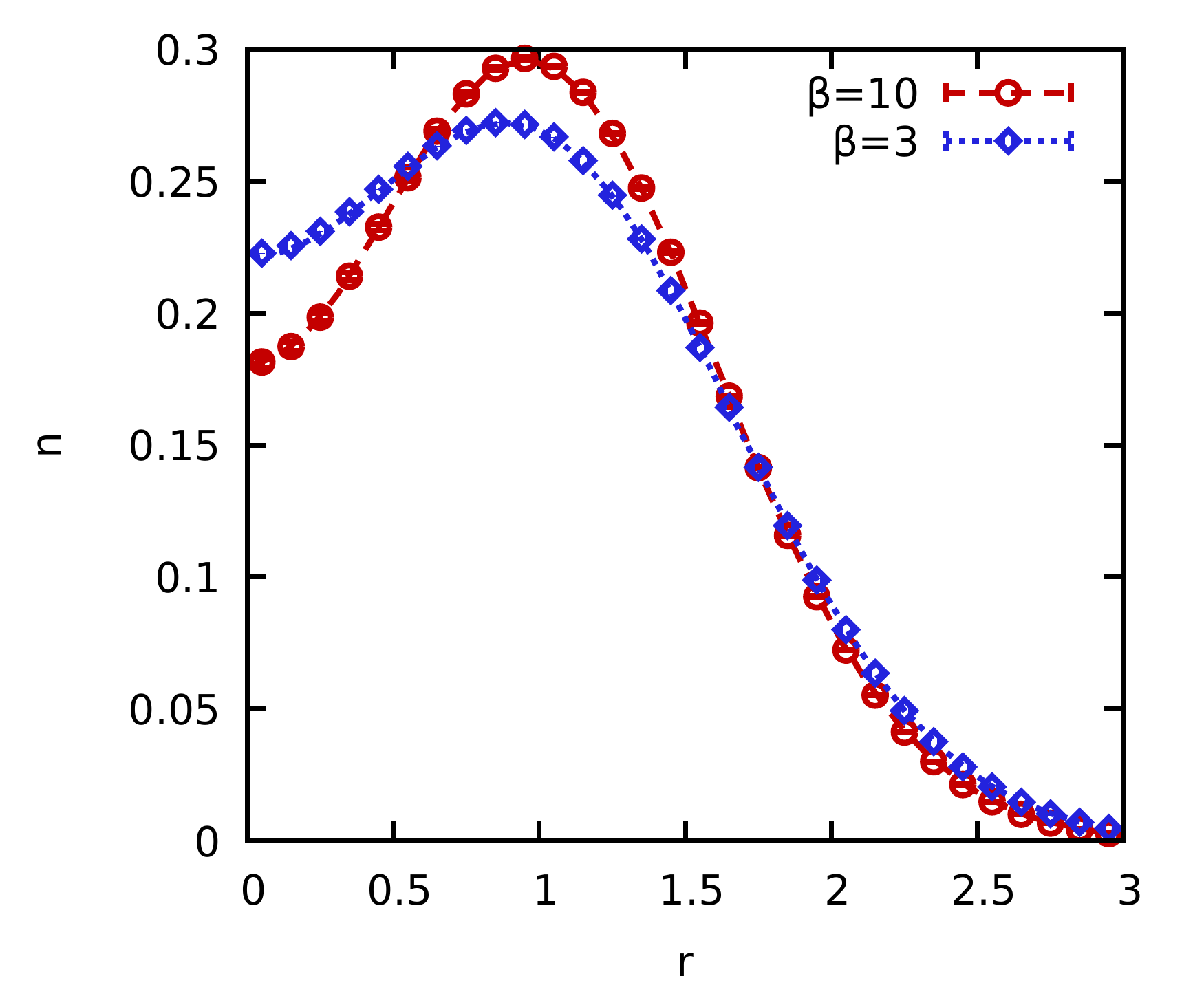}\includegraphics[width=0.5\textwidth]{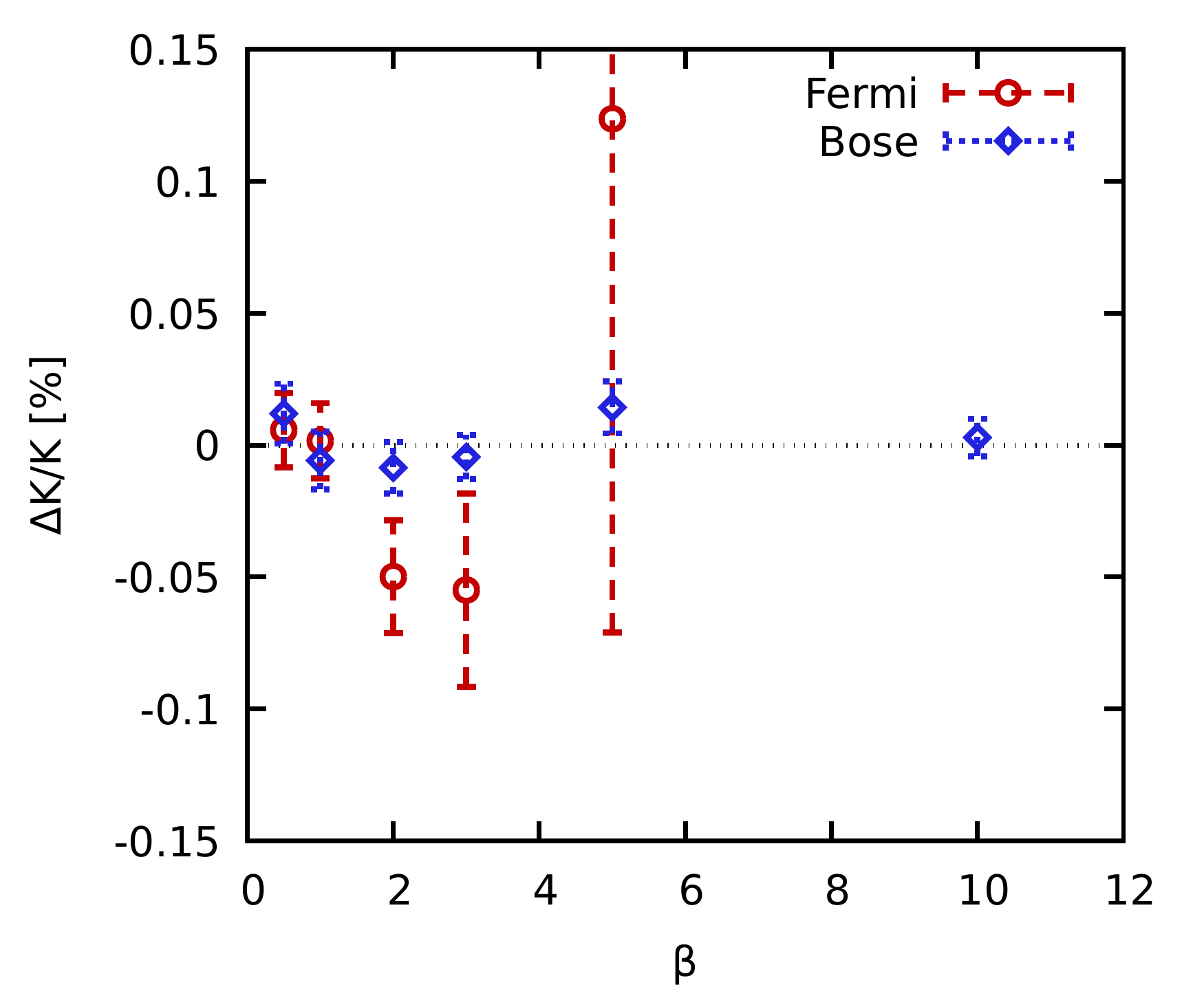}
\caption{\label{fig:virial}
PIMC results for the $\beta$-dependence of electrons in a $2D$ harmonic trap with $\lambda=1.5$ and $\mu=4.4$. Left: Radial density distribution $n(r)$ for $\beta=10$ (red) and $\beta=3$. The points and lines correspond to PIMC data from this work and CPIMC data by Schoof~\cite{Schoof}, respectively. Right: Relative difference (in percent) in the kinetic energy $K$ computed directly from PIMC and from the virial theorem, Eq.~(\ref{eq:virial}). The red circles and blue diamonds have been obtained for Fermi- and Bose-statistics.
}
\end{figure}

Let us start our investigation with a verification of our implementation of the fermionic PIMC method in the GCE. While grand-canonical results for electrons are fairly sparse in the respective literature, a rare example for electrons in a $2D$ harmonic trap has been presented by Schoof~\cite{Schoof} on the basis of the configuration PIMC (CPIMC) method~\cite{Groth_PRB_2016,Schoof_CPP_2015}. The results are shown in the left panel of Fig.~\ref{fig:virial} for $\lambda=1.5$ and $\mu=4.4$ for $\beta=10$ (red) and $\beta=3$ (blue). More specifically, we show the radial density $n(r)$ as a function to the distance to the center of the trap, and the lines and symbols correspond to the CPIMC data and our new PIMC results, respectively. For completeness, we note that the odd value of the chemical potential has been selected to result in an average particle number of $\braket{\hat N}\approx3$ electrons. More importantly, we find perfect agreement between the two independent data sets, which serves as a strong validation of our implementation. From a physical perspective, we find that the interesting interplay between the Coulomb repulsion ($\lambda=1.5$ constitutes an intermediate coupling strength) and the external potential with the fermionic antisymmetry results in a nontrivial density profile with a maximum around $r=1$. Naturally, this peak is more pronounced for the lower temperature, where both correlation and fermionic exchange effects are more pronounced~\cite{Ott2018}.

A further verification of our implementation is depicted in the right panel of Fig.~\ref{fig:virial}. In particular, the well-known virial theorem~\cite{greiner1995thermodynamics} allows us to express the kinetic energy $K$ in terms of the external potential $V_\textnormal{ext}$ and the interaction energy $W$,
\begin{eqnarray}\label{eq:virial}
K = V_\textnormal{ext} - \frac{W}{2}\ .
\end{eqnarray}
The right panel shows the relative difference (in percent) between the straightforward thermodynamic PIMC estimator (see Ref.~\cite{Janke_JCP_1997} for an extensive discussion of different energy estimators in PIMC) for $K$ and the evaluation of Eq.~(\ref{eq:virial}) over the entire relevant range of inverse temperatures $\beta$.
Let us first discuss the blue diamonds that have been obtained for Bose statistics. Evidently, the relative statistical uncertainty remains relatively constant, and the data points fluctuate around zero within the respective error bars. In other words, the two different estimations of the kinetic energy $K$ are equal to within $\sim0.01\%$, and the virial theorem Eq.~(\ref{eq:virial}) is fulfilled by our simulations.
The red circles have been obtained from the same simulations, but were subsequently obtained for Fermi statistics by evaluating Eq.~(\ref{eq:ratio}). These points, too, fluctuate around zero within the given noise level, but the error bars systematically increase with $\beta$. Naturally this is a direct and expected consequence of the FSP, see Eqs.~(\ref{eq:exponential}) and (\ref{eq:MC_error}) above.

\begin{figure}\centering
\includegraphics[width=0.5\textwidth]{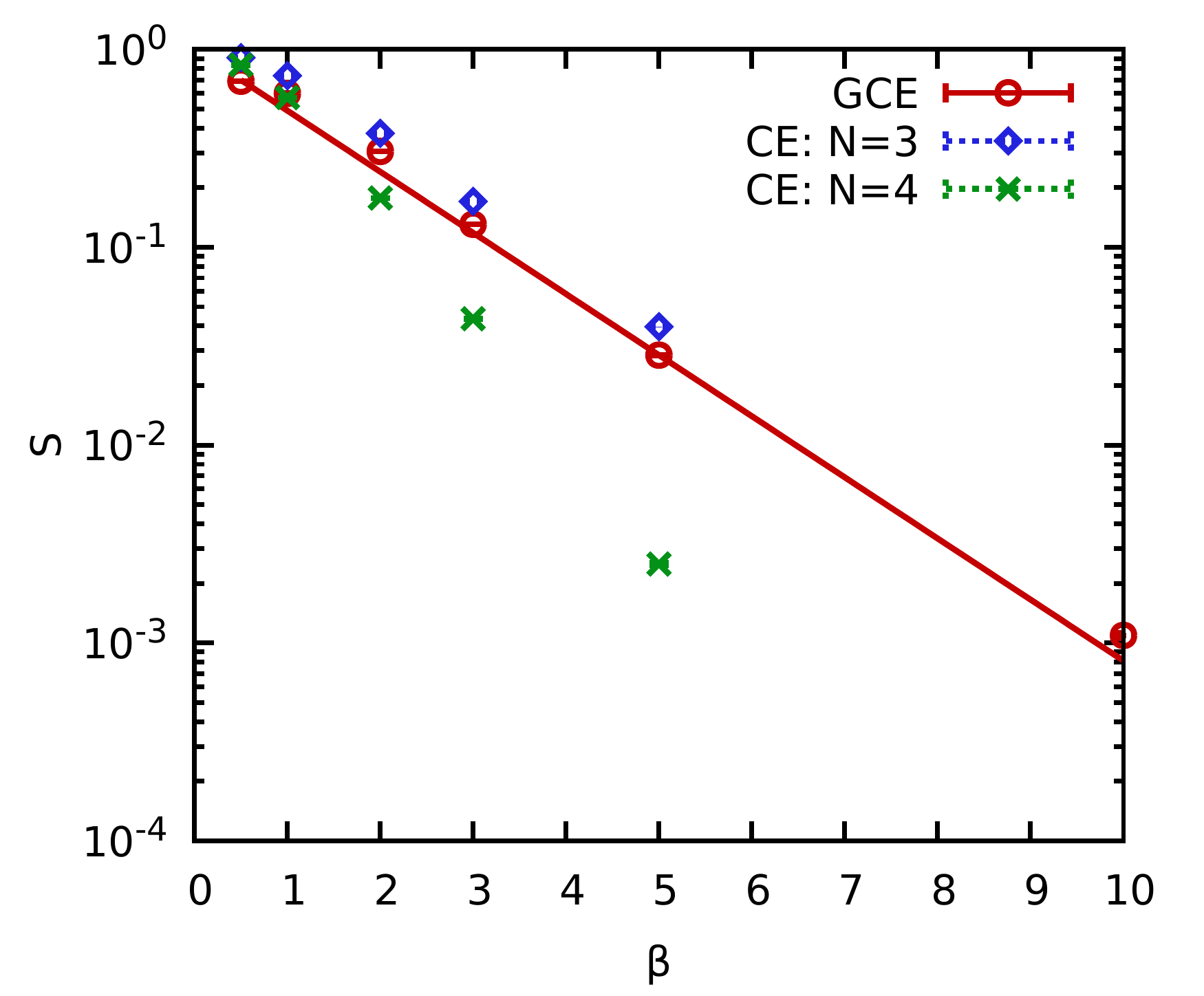}\includegraphics[width=0.5\textwidth]{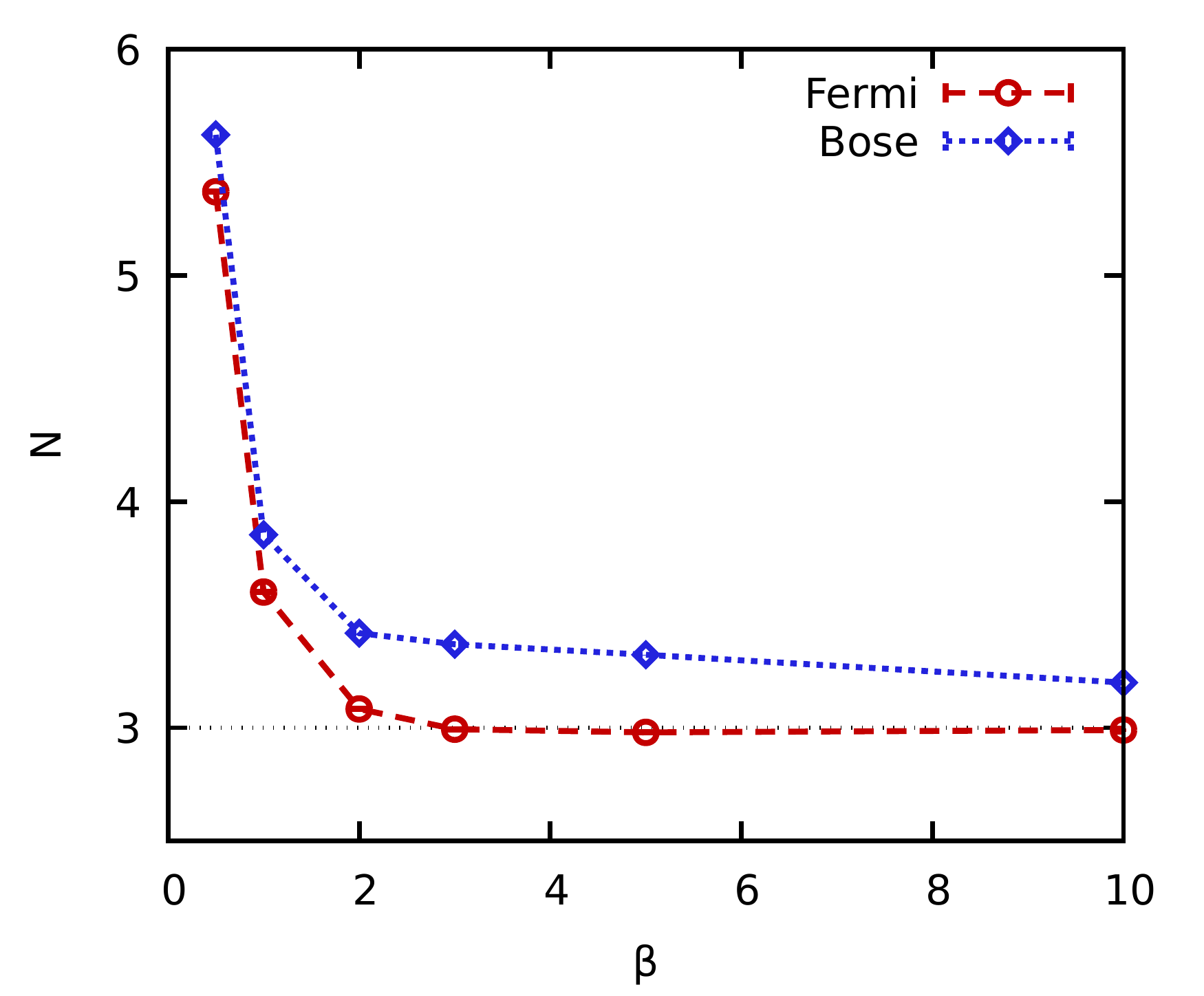}
\caption{\label{fig:sign}
PIMC results for the $\beta$-dependence of electrons in a $2D$ harmonic trap with $\lambda=1.5$ and $\mu=4.4$. Left: Average sign $S$, red circles: GCE; blue diamonds: CE with $N=3$; green crosses: CE with $N=4$. Right: average particle number $\braket{\hat N}$ in the GCE for Fermi- (red circles) and Bose-statistics (blue diamonds).
}
\end{figure}

Let us next proceed to the central topic to be discussed in this work, i.e., the FSP itself. To this end, we show the $\beta$-dependence of $S$ in the left panel of Fig.~\ref{fig:sign} for the same conditions as in Fig.~\ref{fig:virial}. The red circles show results for the GCE, and the blue diamonds and green crosses have been obtained in the CE for $N=3$ and $N=4$. Evidently, all three data sets exhibit the expected monotonous decrease with $\beta$ and attain unity in the limit of $\beta\to0$ (i.e., $T\to\infty$) when quantum degeneracy effects completely vanish. For the two highest temperatures, the grand-canonical simulations actually exhibit the smallest values of $S$, whereas it remains between the other curves for larger $\beta$.
This can be readily explained by the PIMC results for the average particle number shown in the right panel of Fig.~\ref{fig:sign}, where the red circles and blue diamonds distinguish data for Fermi- and Bose-statistics. 
For $\beta=0.5$ and $\beta=1$, the fermionic results for $\braket{\hat N}$ are not yet converged to the ground-state result, whereas it remains relatively constant for $\beta\gtrsim2$. Naturally, the GCE result for $S$ at $\beta=0.5$ is smaller than the CE results for both $N=3$ and $N=4$, when the corresponding average particle number exceeds $N=5$.

Interestingly, the bosonic expectation value for $\braket{\hat N}$ exhibits a substantially slower convergence with $\beta$, and has not yet converged even for $\beta=10$. Heuristically, this can be understood as follows: for fermions, the Pauli exclusion principle prevents the multiple occupation of the same state, such that each additional particle results in a relatively large increase in the energy; bosons, on the other hand, can potentially all occupy the same state (this is associated with Bose-Einstein condensation~\cite{landau}), and the corresponding increase in the energy for an additional particle is, on average, much lower. Hence, bosons exhibit both larger particle number fluctuations and a slower convergence of $\braket{\hat N}$ for low temperature.

Finally, we note that the decay of the grand-canonical sign with $\beta$ is not quite exponential; see the deviations to the solid red line that has been obtained from a fit of the form
\begin{eqnarray}
S(\beta)=a e^{-b\beta}\ ,
\end{eqnarray}
with $a$ and $b$ being the free parameters. This, too, is a direct consequence of the different behaviour of $\braket{\hat N}$ for bosons and fermions. More specifically, the average sign in the GCE is consistently lower than the canonical sign for $N=3$, even though it holds $\braket{\hat N}\approx3$. Yet, as we have seen in Sec.~\ref{sec:GCE} above, the grand-canonical sign is given by the superposition of the respective canonical signs $\braket{\hat S}_N$, but weighted with the bosonic probability of that particle-number sector $P(N)=N_N/N_\textnormal{MC}$. Therefore, configurations with, say, $N=4$ affect the grand-canonical sign at $\beta=10$, even though these configurations do not contribute to the expectation values of physical observables like the energy. Indeed, $P(4)$ vanishes for fermions in this case within the Monte Carlo error bars, whereas it is $P(4)\approx0.2$ for bosons.

In this sense, a fermionic PIMC simulation in the GCE at low temperature will reproduce the expectation values in the CE with the same $N$, but with a more severe sign problem as the bosonic reference system has a broader distribution $P(N)$.

\begin{figure}\centering
\includegraphics[width=0.85\textwidth]{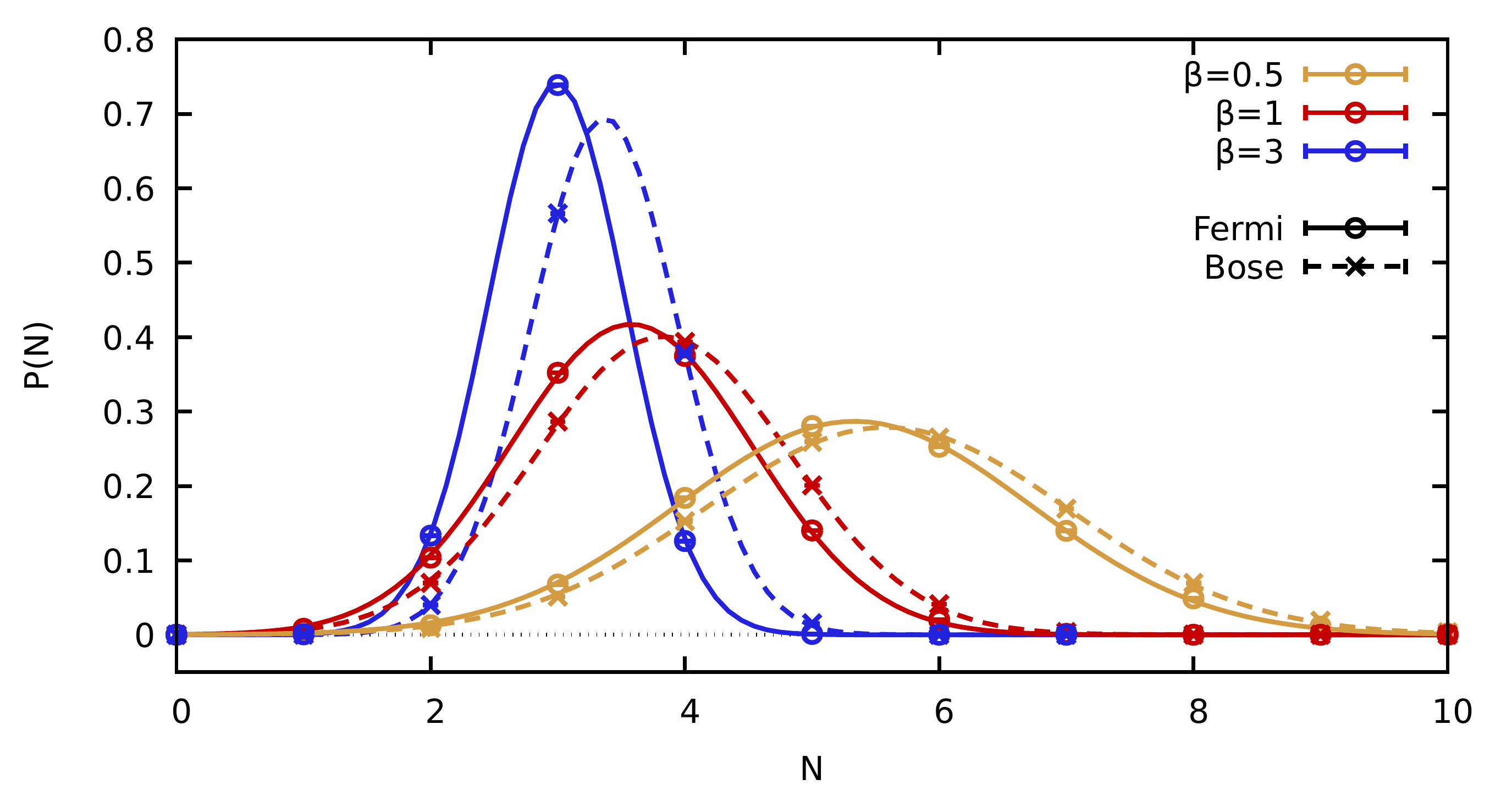}
\caption{\label{fig:histogram}
Histograms of particle numbers $N$ within a grand-canonical PIMC simulation of electrons in a $2D$ harmonic trap with $\lambda=1.5$ and $\mu=4.4$. The circles (solid lines) and crosses (dashed lines) show PIMC data (Gaussian fits) for Fermi- and Bose-statistics. The yellow, red, and blue curves have been obtained for $\beta=0.5$, $\beta=1$, and $\beta=3$, respectively. 
}
\end{figure}

Let us conclude our investigation of the temperature dependence of the FSP in the GCE by looking at the distributions of particle numbers $P(N)$. This is shown in Fig.~\ref{fig:histogram}, where the circles and crosses distinguish PIMC results for Fermi- and Bose-statistics, and the yellow, red, and blue curves have been obtained for $\beta=0.5$, $\beta=1$, and $\beta=3$, respectively. In addition, the lines depict Gaussian fits to the PIMC data, i.e.,
\begin{eqnarray}\label{eq:Gauss}
P(N) = \frac{\textnormal{exp}\left(-\frac{(N-\mu)^2}{2\sigma^2}\right)}{\sqrt{2\pi\sigma^2}}\ ,
\end{eqnarray}
with $\sigma$ and $\mu$ being the two free parameters.
We note the excellent agreement between the data and Eq.~(\ref{eq:Gauss}) for all shown cases. Furthermore, we find that the bosonic distribution is always shifted to larger $N$ compared to the case of fermions, and this shift increases towards lower temperature. As mentioned above, this is one of the central reasons for the more severe sign problem in the GCE compared to the CE investigated in Ref.~\cite{dornheim_sign_problem}. Finally, the particle number fluctuation increases with $T$, as it is expected.

\subsection{Dependence on the chemical potential\label{sec:mu}}

\begin{figure}\centering
\includegraphics[width=0.5\textwidth]{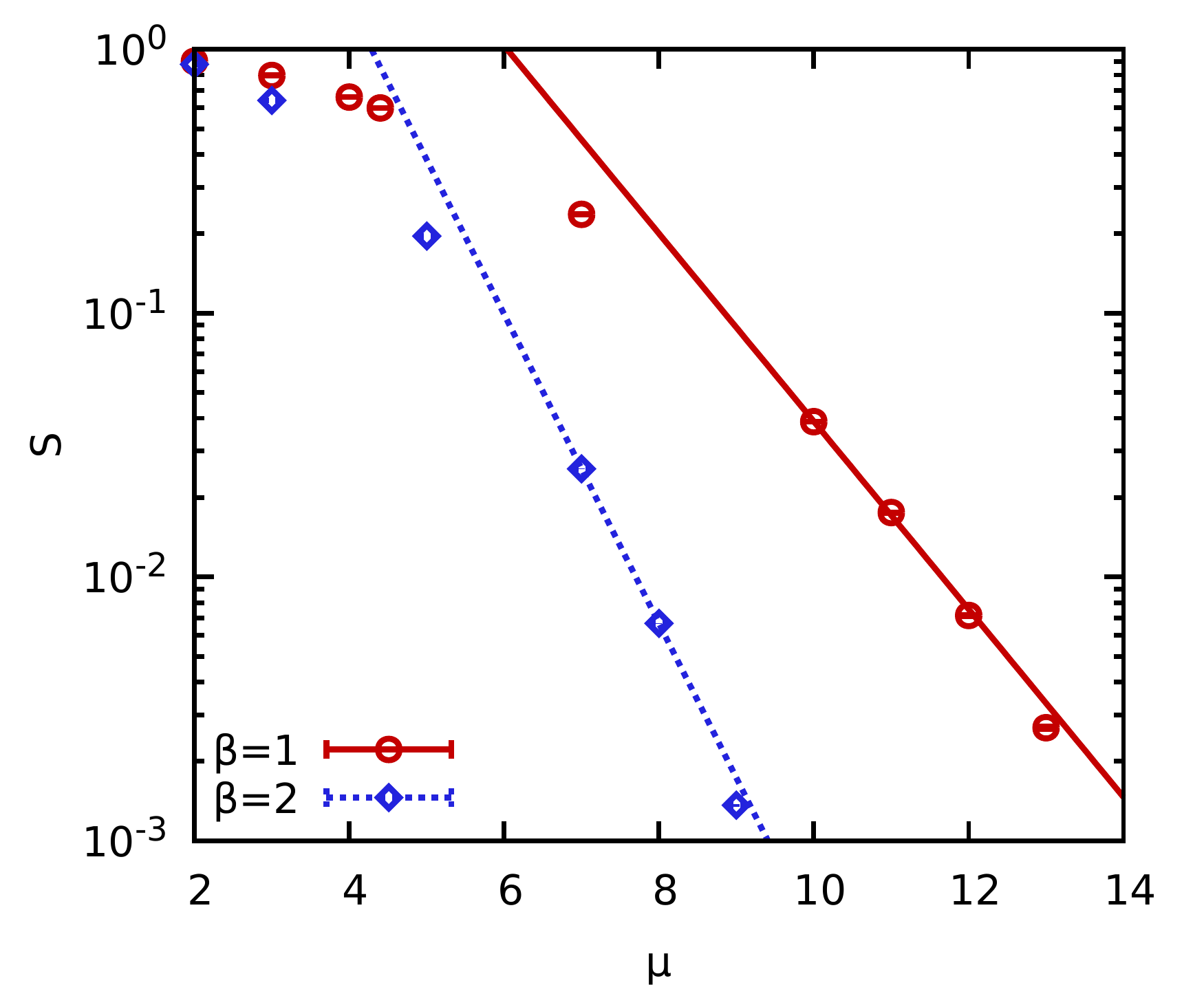}\includegraphics[width=0.5\textwidth]{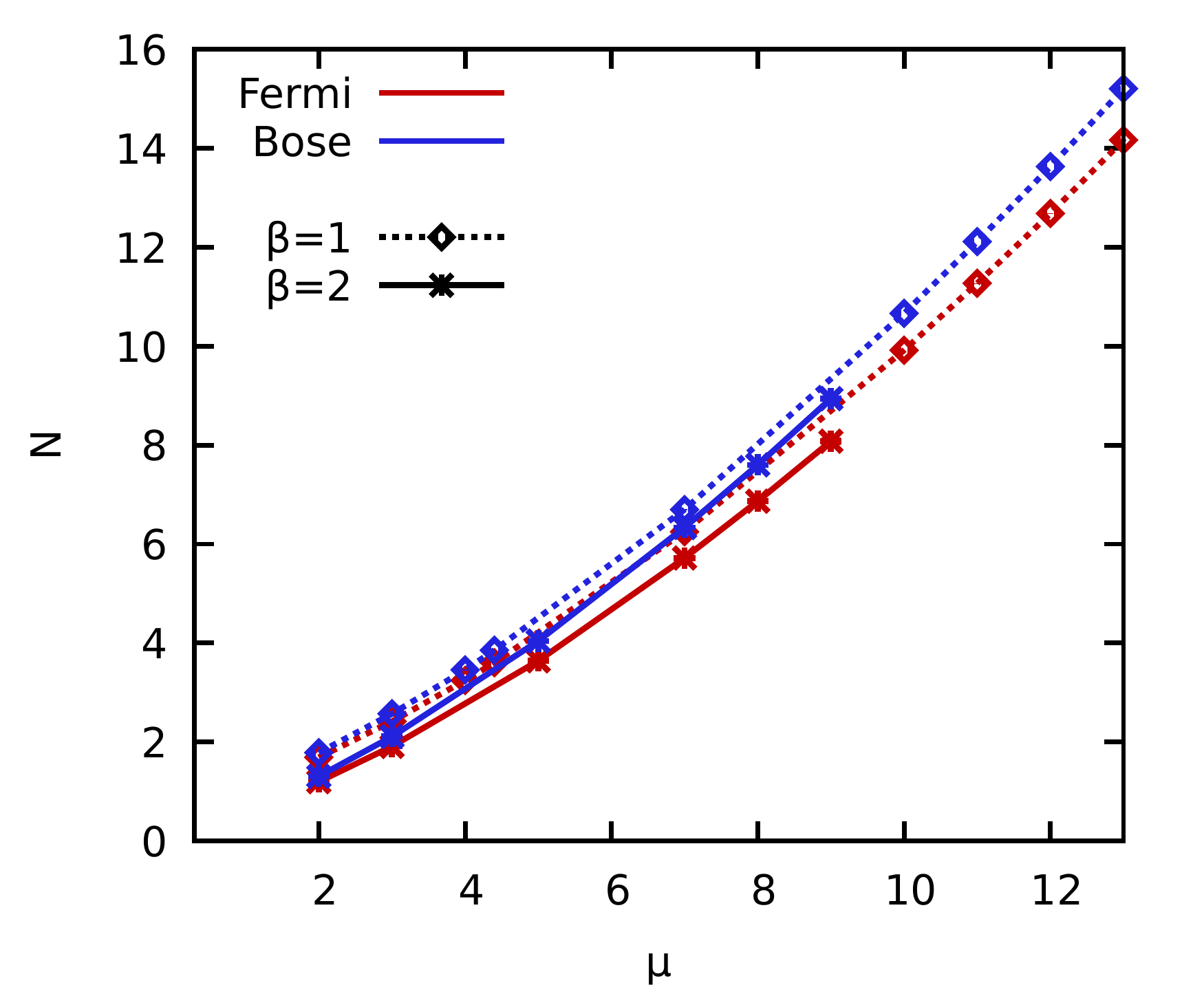}
\caption{\label{fig:sign_mu}
PIMC results for the $\mu$-dependence of electrons in a $2D$ harmonic trap with $\beta=1$ and $\lambda=1.5$. Left: Average sign $S$ in the GCE for $\beta=1$ (red circles) and $\beta=2$ (blue diamonds). Right: Corresponding average particle numbers, with red and blue lines distinguishing bosons and fermions, and dotted diamonds and solid stars corresponding to $\beta=1$ and $\beta=2$.
}
\end{figure}

Let us next investigate the dependence of the FSP in the GCE on the chemical potential $\mu$ shown in Fig.~\ref{fig:sign_mu}. The left panel shows our PIMC results for the average sign $N$ for $\lambda=1.5$, with the red circles (blue diamonds) corresponding to $\beta=1$ ($\beta=2$). For both temperatures, we observe a monotonous and fast decrease of $S$ upon increasing $\mu$, as it is expected. More specifically, an increase in the chemical potential leads to a larger average particle number. This is shown in the right panel of Fig.~\ref{fig:sign_mu}, where the red and blue curves distinguish bosons and fermions, and the diamonds and crosses have been obtained for $\beta=1$ and $\beta=2$. Indeed, we find that $N(\mu)$ exhibits a monotonous increase that is somewhat faster than linear. 

Conversely, the average sign itself decreases faster than exponential with $N$. This can be directly seen by comparing to the dotted and solid straight lines, which have been obtained from exponential fits of the form
\begin{eqnarray}
S(\mu)=a_\mu e^{-\mu b_\mu}\ ,
\end{eqnarray}
over certain finite $\mu$-intervals, with $a_\mu$ and $b_\mu$ being the free parameters.

\begin{figure}\centering
\includegraphics[width=0.85\textwidth]{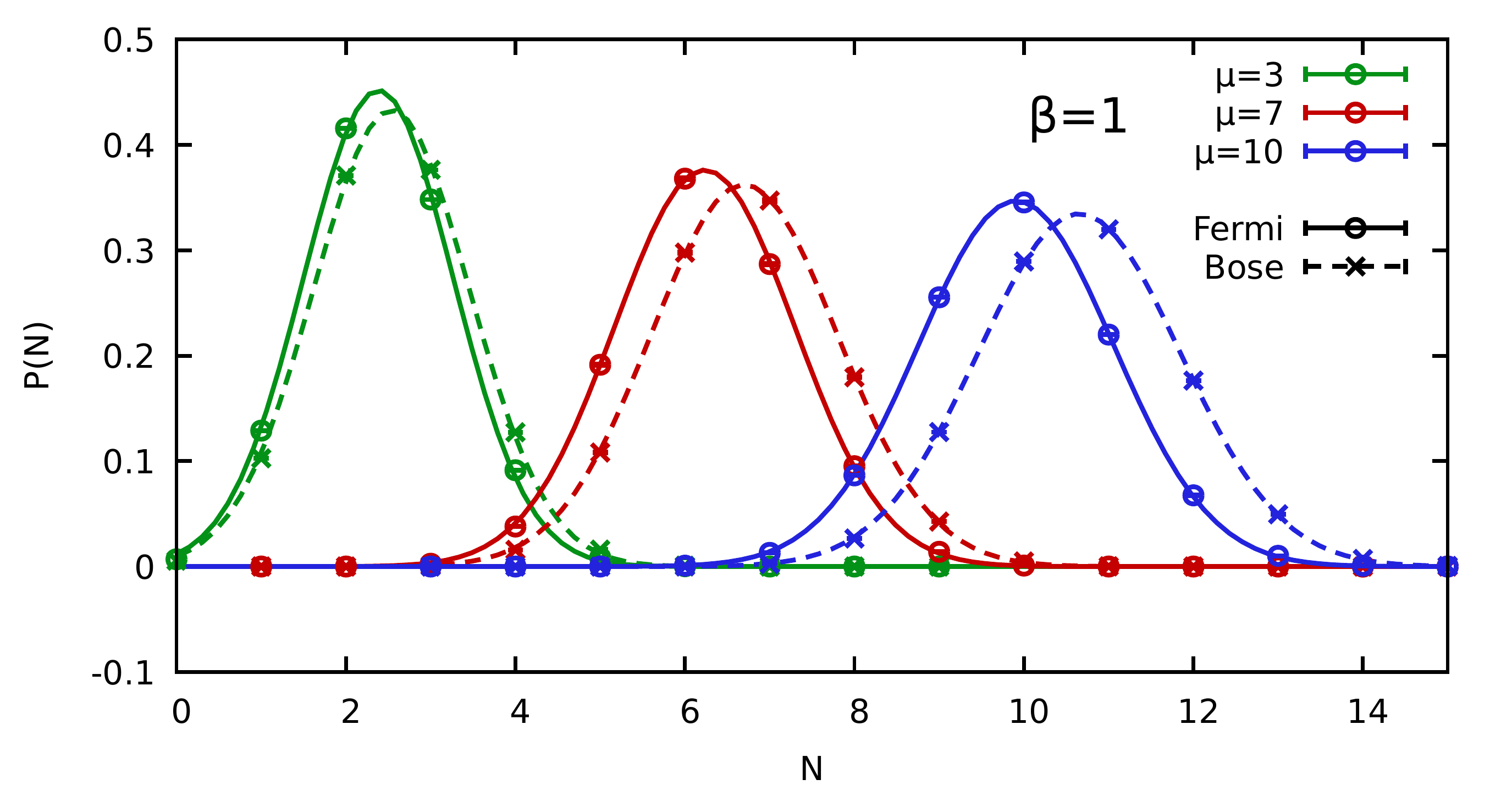}\\\vspace*{-0.84cm}
\includegraphics[width=0.85\textwidth]{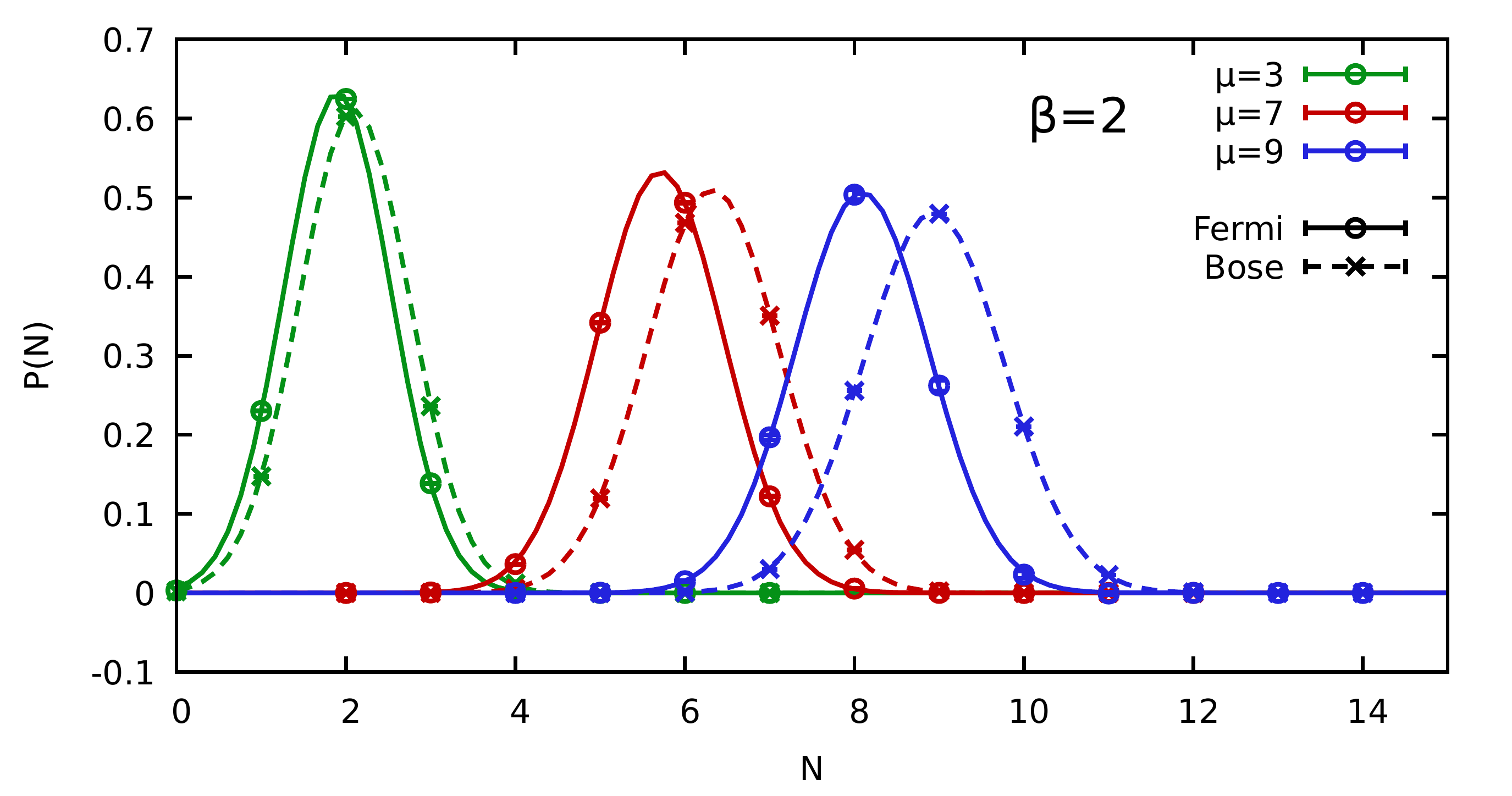}
\caption{\label{fig:histogram_mu}
Histograms of particle numbers $N$ within a grand-canonical PIMC simulation of electrons in a $2D$ harmonic trap with $\lambda=1.5$ and $\beta=1$ (top) and $\beta=2$ (bottom). The circles (solid lines) and crosses (dashed lines) show PIMC data (Gaussian fits) for Fermi- and Bose-statistics for different values of the chemical potential $\mu$.
}
\end{figure}

An additional degree of freedom in the GCE that can potentially lead to a super exponential decrease of $S$ is the fluctuation of $\braket{\hat N}$, i.e., the distribution $P(N)$. The latter is shown in Fig.~\ref{fig:histogram_mu} both for $\beta=1$ (top panel) and $\beta=2$ (bottom panel). For both temperatures, the general trend is the same: upon increasing the chemical potential, the mean values of the Gaussian distributions are shifted to larger $N$, as it is expected. In addition, larger values of $\mu$ result in a broader distribution, although this trend is relatively small for the chosen parameters. More important, larger chemical potentials also lead to more pronounced deviations between the bosonic and fermionic distributions, which further contributes to the drastic decrease of the sign observed in Fig.~\ref{fig:sign_mu}.

Finally, we find that the distributions $P(N)$ are more peaked for the lower temperature, which is consistent to the behaviour observed in Sec.~\ref{sec:temperature} above.

\subsection{Dependence on the coupling strength\label{sec:coupling}}

\begin{figure}\centering
\includegraphics[width=0.5\textwidth]{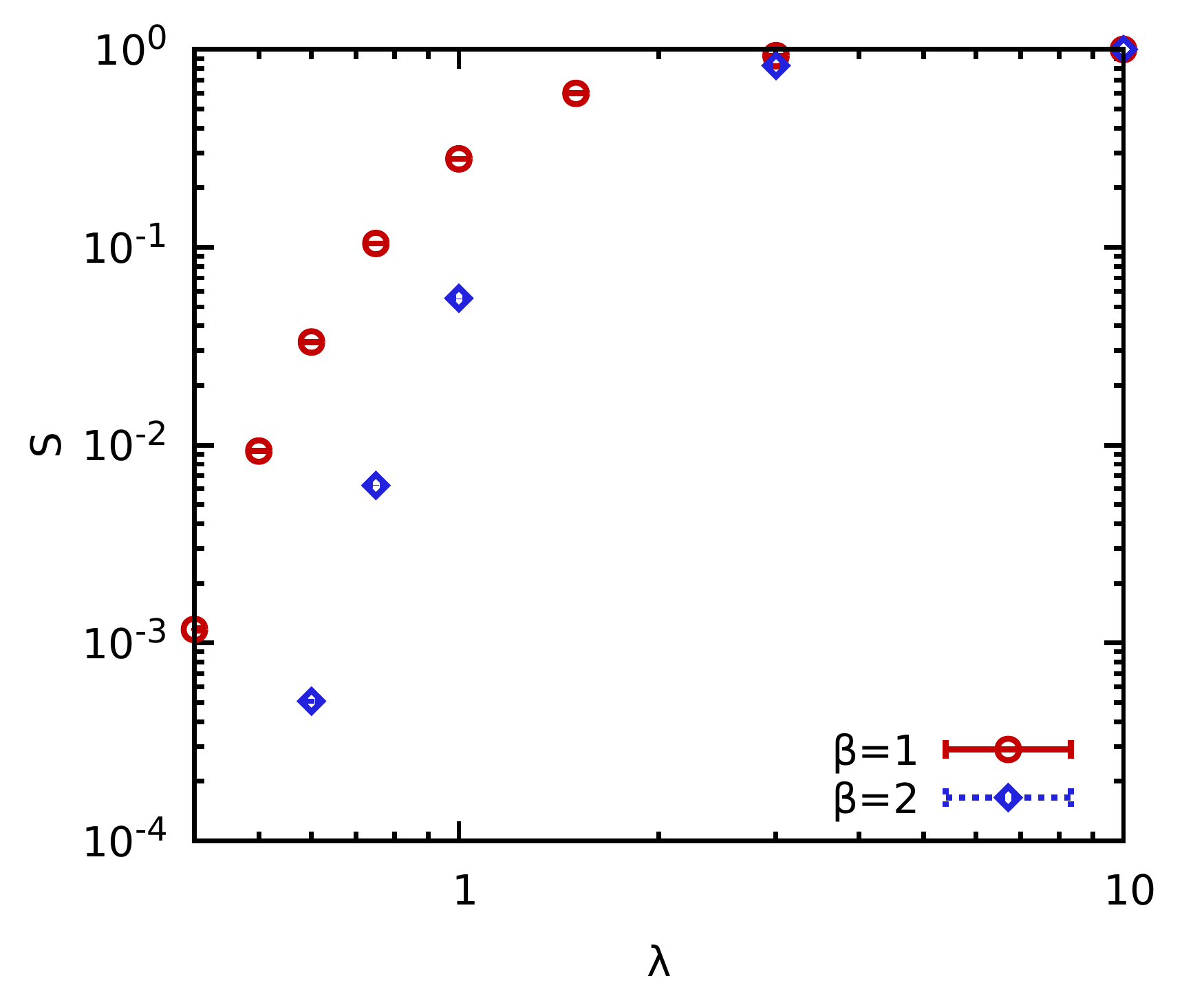}\includegraphics[width=0.5\textwidth]{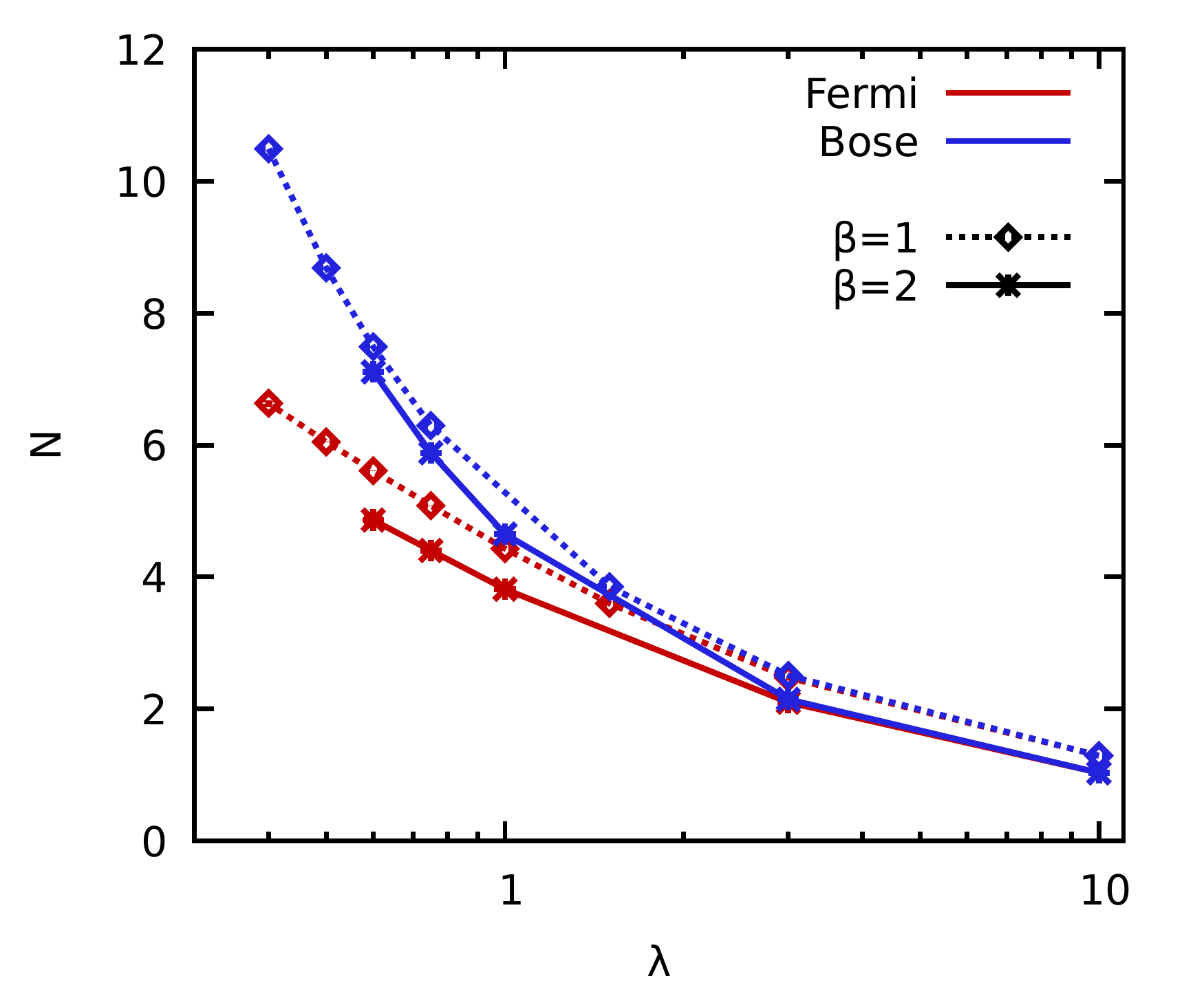}
\caption{\label{fig:sign_lambda}
PIMC results for the $\lambda$-dependence of electrons in a $2D$ harmonic trap with $\mu=4.4$. Left: Average sign $S$ for $\beta=1$ (red circles) and $\beta=2$ (blue diamonds). Right: Average particle number $\braket{\hat N}$ for bosons (blue) and fermions (red), with the diamonds and crosses distinguishing $\beta=1$ and $\beta=2$.
}
\end{figure}

The final variable to be investigated in the present work is the dependence of the FSP in the GCE on the coupling parameter $\lambda$.
This is investigated in the left panel of  Fig.~\ref{fig:sign_lambda}, where we show the $\lambda$-dependence of $S$ for $\mu=4.4$ with $\beta=1$ (red circles) and $\beta=2$ (blue diamonds). In particular, both data sets exhibit the same overall trend: for large values of $\lambda$, the paths of individual particles are effectively separated by the strong Coulomb repulsion. Consequently, quantum exchange effects are negligible and the sign eventually attains unity. With decreasing $\lambda$, the paths begin to overlap, permutation-cycles appear within the PIMC simulation with increasing frequency and the average sign $S$ drops. Indeed, it is easy to see from Fig.~\ref{fig:sign_lambda} that this drop is super-exponential for the depicted parameters. 

To understand this finding, we again consider the average number of particles $\braket{\hat N}$, which we show in the right panel of the same figure. Here the red curves have been obtained for Fermi- and the blue curves for Bose-statistics, and the diamonds and crosses distinguish $\beta=1$ and $\beta=2$. For large values of $\lambda$, the bosonic and fermionic curves are identical, whereas they diverge as the system becomes more ideal. Most strikingly, the bosonic curves exhibit a substantially larger increase of $\braket{\hat N}$ with decreasing $\lambda$ compared to the fermionic curves. From a physical perspective, this is again a direct consequence of the Pauli exclusion principle, which prevents the multiple occupation of energetically low lying states in the case of Fermi-statistics.
At the same time, this divergence in $\braket{\hat N}$ between the two types of particles explains the dramatic drop in $S$.

\begin{figure}\centering
\includegraphics[width=0.85\textwidth]{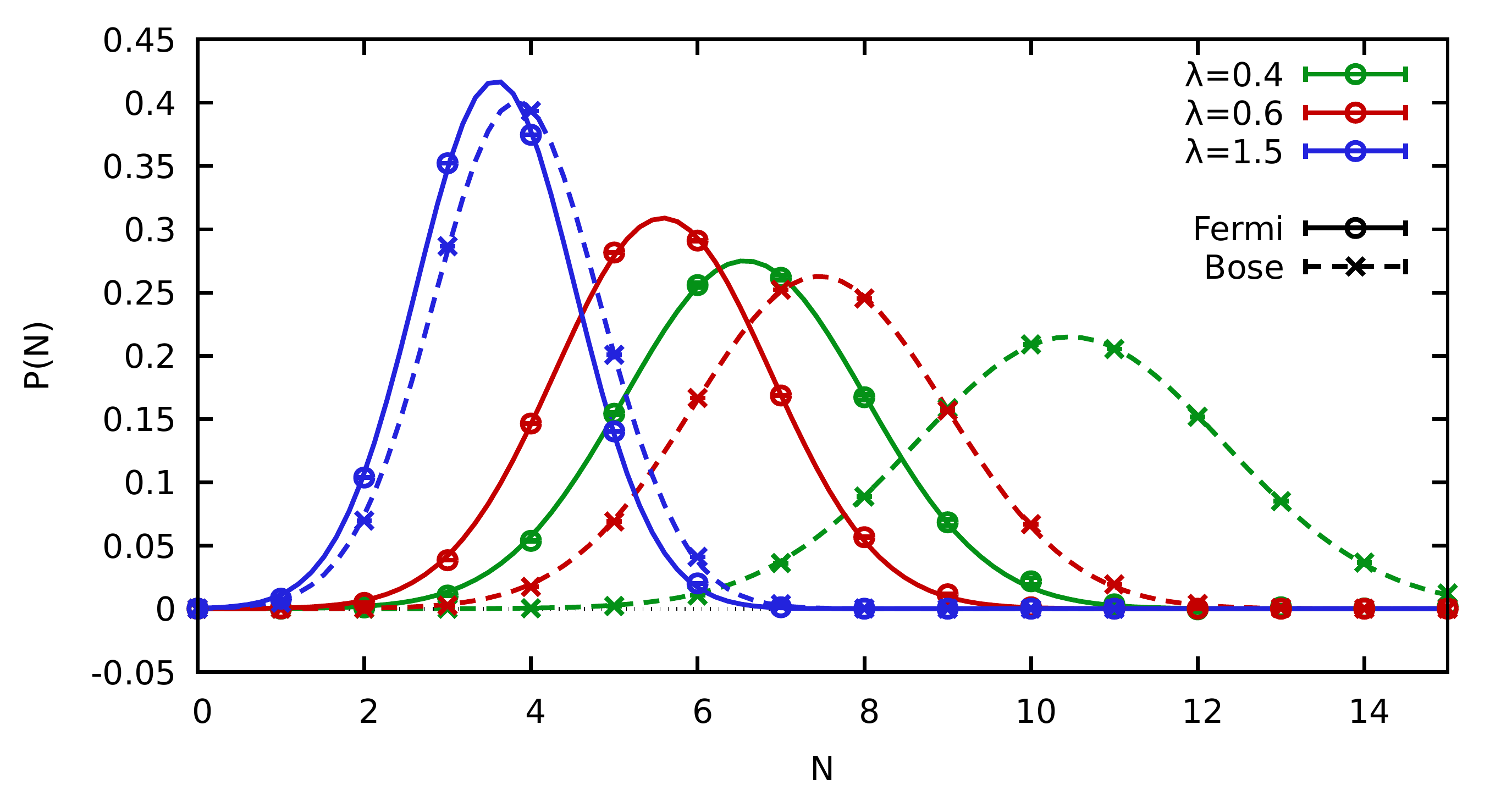}
\caption{\label{fig:histogram_lambda}
Histograms of particle numbers $N$ within a grand-canonical PIMC simulation of electrons in a $2D$ harmonic trap with $\beta=1$ and $\mu=4.4$. The circles (solid lines) and crosses (dashed lines) show PIMC data (Gaussian fits) for Fermi- and Bose-statistics. 
}
\end{figure}

Let us conclude this section by considering the particle number distribution $P(N)$ itself, which we show in Fig.~\ref{fig:histogram_lambda} for $\beta=1$ for three different values of $\lambda$. As usual the solid and dashed lines show Gaussian fits [cf.~Eq.~(\ref{eq:Gauss})] for fermions and bosons, and the circles and crosses depict the corresponding PIMC data. 

For $\lambda=1.5$ (blue), the system is moderately coupled and quantum exchange-effects play a relatively small role at this value of $\beta$. Consequently, the bosonic and fermionic distributions are quite similar both in $\sigma$ and $\mu$.
For $\lambda=0.6$ (red), the coupling strength is no longer the dominant effect, and both curves are shifted towards significantly larger values of $N$, and the distributions are broader. Furthermore, there appears a substantial difference between the results for fermions and bosons both in the position and the width. 
Finally, the green curves have been obtained for an even more weakly coupled system with $\lambda=0.4$. For fermions, this leads to a slightly shifted, slightly broader curve compared to $\lambda=0.6$. In stark contrast, the corresponding bosonic distribution function is shifted to drastically larger values of $N$.

This plainly illustrates a crucial bottleneck of fermionic PIMC simulations in the GCE, which is absent in the CE: the actual bosonic simulation may potentially spend most time in configurations with particle numbers that do not contribute to the fermionic grand-canonical expectation value of interest. In the case of $\lambda=0.4$, nearly half the simulation time is spent in configurations with $N\geq11$, for which the fermionic $P(N)$ is close to zero.  In addition, configurations with $N=4$ substantially contribute to the Fermi system, but this sector is hardly visited within the effectively bosonic PIMC simulation.

We thus conclude that, all other parameters being equal, grand-canonical simulations are generally afflicted with an even more severe sign problem compared to simulations in the canonical ensemble. Therefore, they should only be undertaken if this is necessary to obtain a desirable physical property of interest such as the compressibility or the Matsubara Green function.

\subsection{Distribution of expectation values\label{sec:distribution}}

\begin{figure}\centering
\includegraphics[width=0.485\textwidth]{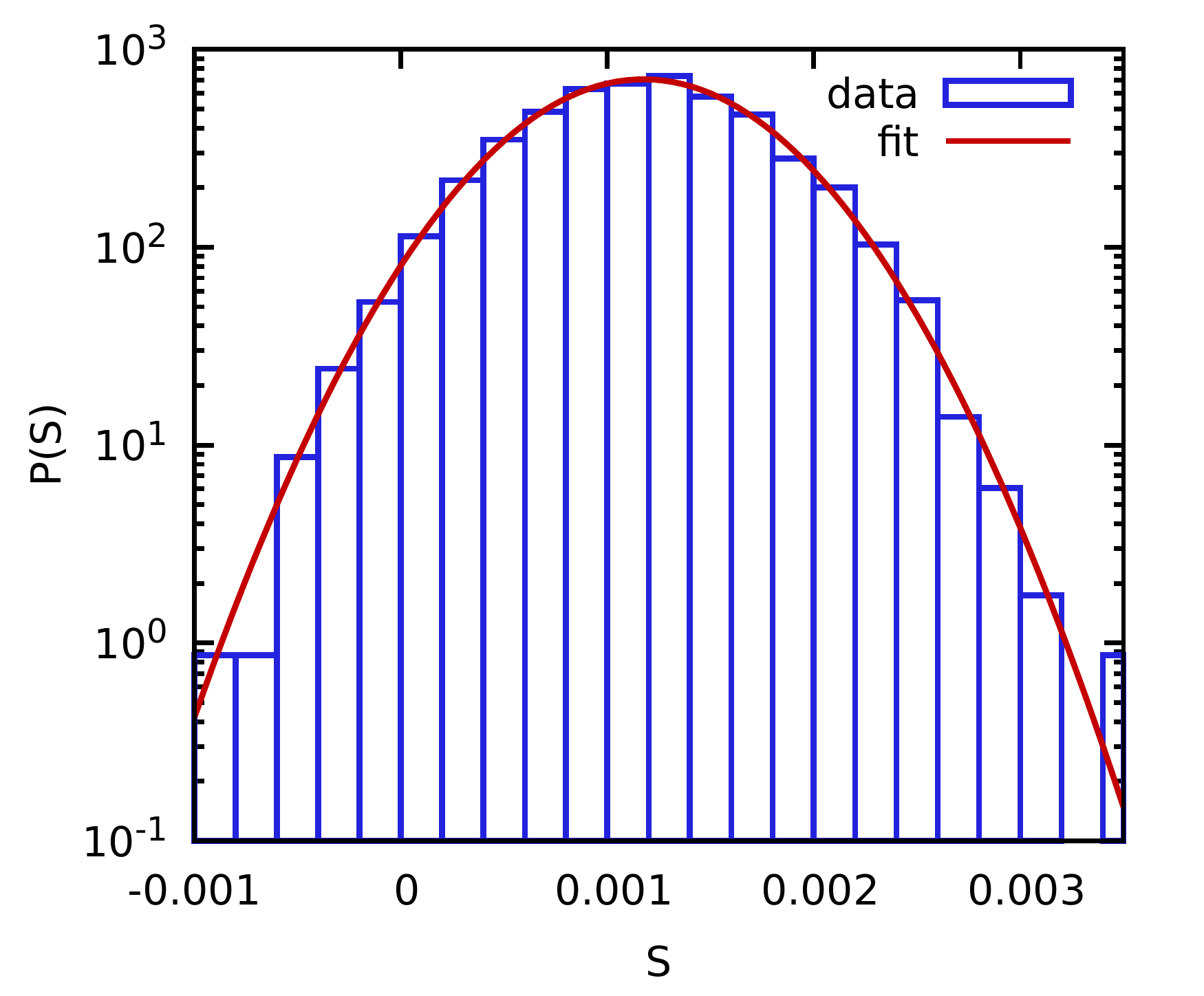}\includegraphics[width=0.485\textwidth]{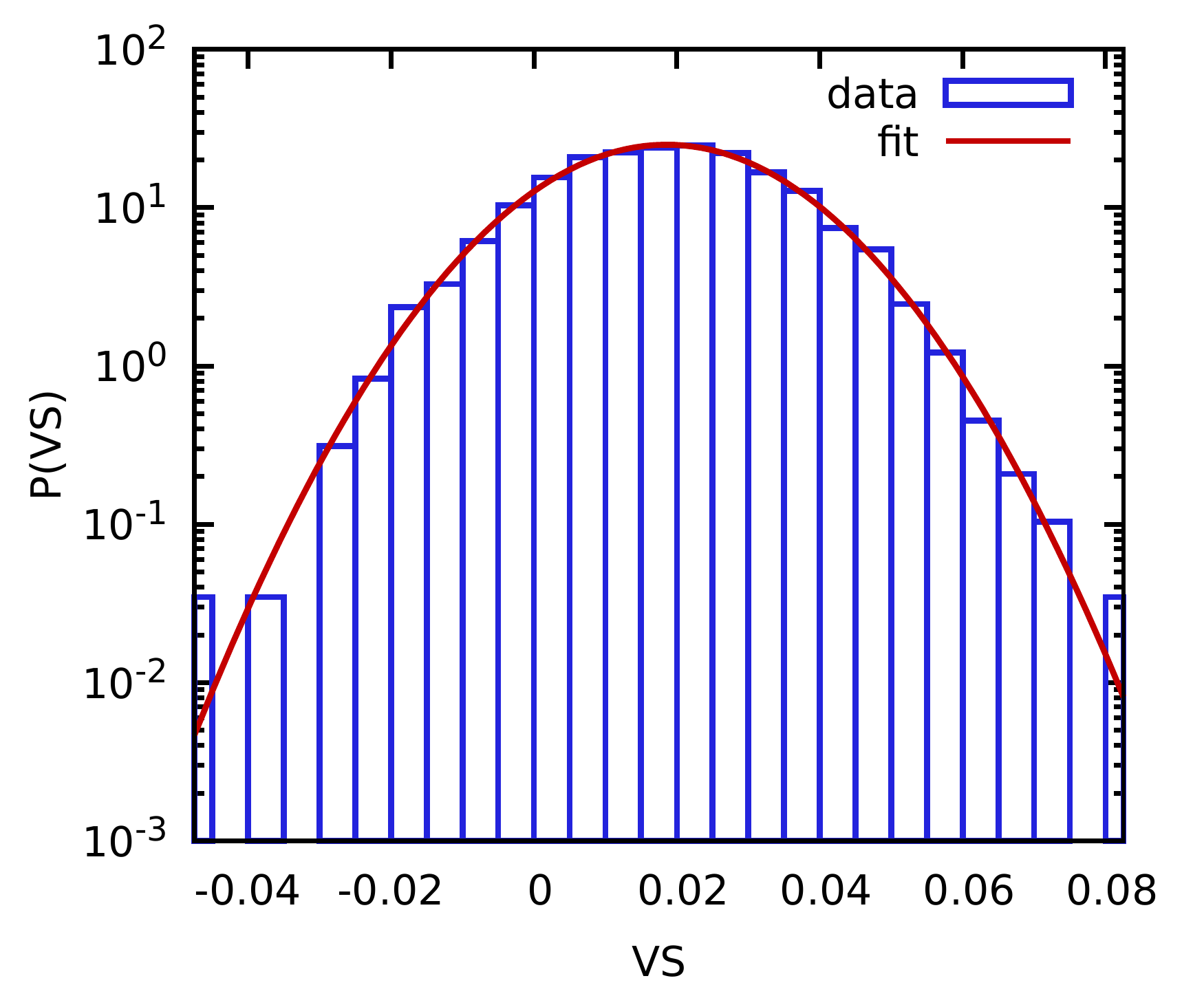}
\caption{\label{fig:expectation}
Histograms of expectation values from $N\approx6000$ independent PIMC simulations with $N_\textnormal{MC}\approx4.8\times10^7$ measurements per simulation for a system with $\lambda=0.4$, $\mu=4.4$, and $\beta=1$. The left and right panels correspond to the denominator and enumerator of Eq.~(\ref{eq:ratio}) for the case of $\hat O$ being the total potential energy $V$.
}
\end{figure}

Let us conclude our investigation of the FSP in the GCE by considering the distribution of the different expectation values. As we have already mentioned in Sec.~\ref{sec:FSP} above, the Monte Carlo estimate of the expectation value $\braket{\hat O}'$ is a random variable that is normally distributed around the exact result. To verify this prediction by the central limiting theorem, we have performed $M\approx6000$ independent PIMC simulations of the system with $\lambda=0.4$, $\mu=4.4$, and $\beta=1$ with $N_\textnormal{MC}\approx4.8\times10^7$ measurements each. In the left (right) panel of Fig.~\ref{fig:expectation}, we show the corresponding histogram of $\braket{\hat S}'$ ($\braket{\hat V\hat S}'$), i.e., the denominator (enumerator) of Eq.~(\ref{eq:ratio}) for the total potential energy $V$. More specifically, the blue bars depict the histogram of PIMC data and the solid red line Gaussian fits according to Eq.~(\ref{eq:Gauss}). Evidently, the fitted curves are in excellent agreement to the data, and the simulations are quasi-exact.

\begin{figure}\centering
\includegraphics[width=0.485\textwidth]{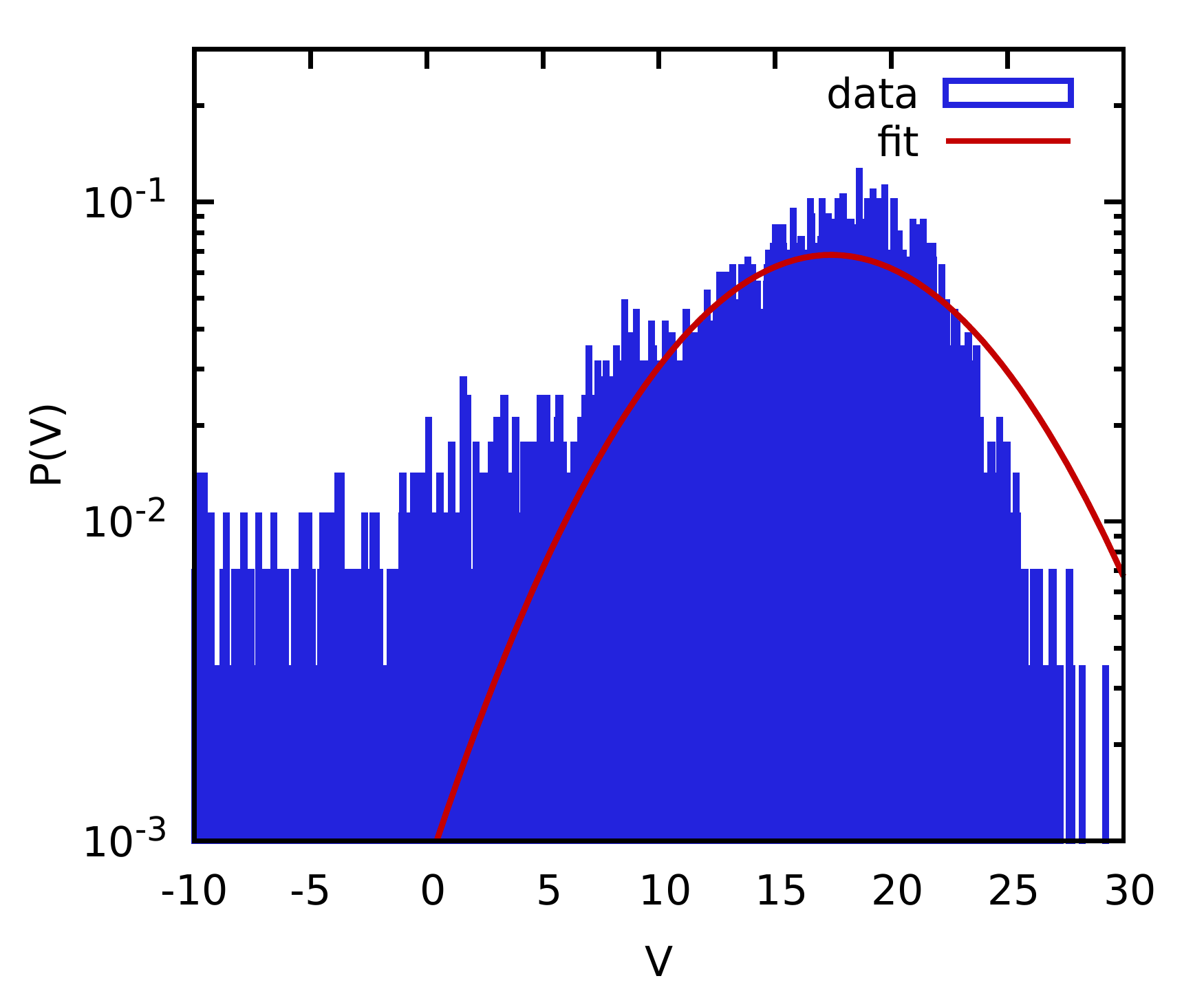}\includegraphics[width=0.485\textwidth]{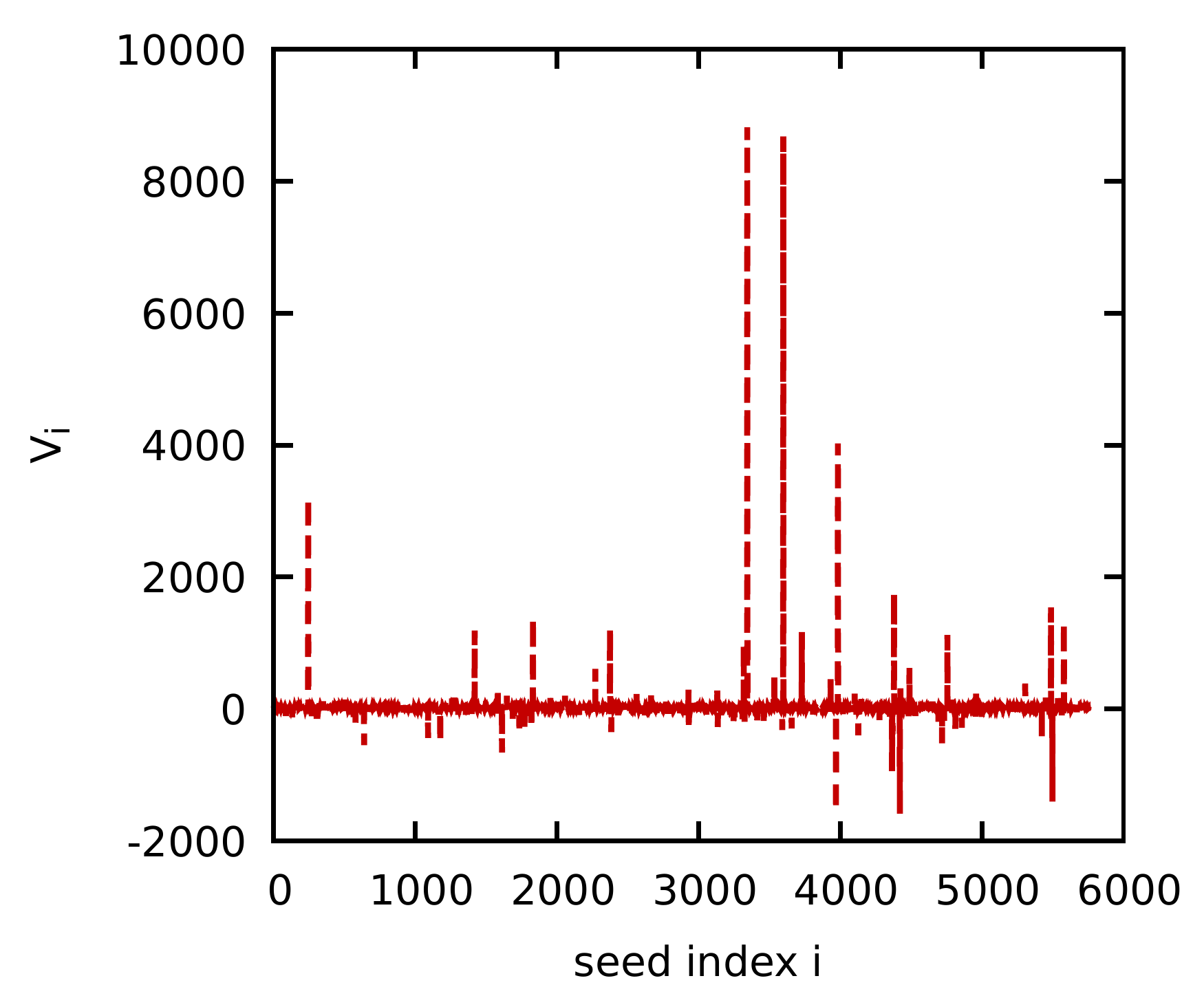}
\caption{\label{fig:expectation2}
Histograms of expectation values from $N\approx6000$ independent PIMC simulations with $N_\textnormal{MC}\approx4.8\times10^7$ measurements per simulation for a system with $\lambda=0.4$, $\mu=4.4$, and $\beta=1$. The left panel shows the distribution of the evaluation of the ratio from Eq.~(\ref{eq:ratio}) for each individual PIMC simulation, and the right panel the corresponding results for all of them.
}
\end{figure}

Yet, we are not interested in either $\braket{\hat V\hat S}'$
or $\braket{\hat S}'$ for their own sake, but instead require the evaluation of their ratio. The corresponding evaluation of Eq.~(\ref{eq:ratio}) for each individual PIMC simulation leads to the histogram shown in the left panel of Fig.~\ref{fig:expectation2}.
Evidently, the distribution of the ratio is not given by a simple normal distribution, and the fit (solid red) does neither qualitatively nor quantitatively reproduce the obtained data. In particular, there appear individual results for $V$ that are clearly outside of the usual $3\sigma$-interval. This can be seen particularly well in the right panel of the same figure, where we show the results of Eq.~(\ref{eq:ratio}) for all independent PIMC simulations. While most $V_i$ are distributed around the mean value $V\sim10$, there appear numerous drastic outliers with maximum values exceeding $V_i=8000$.

This can be understood in the following way: going back to histogram of $\braket{\hat S}$ shown in the left panel of Fig.~\ref{fig:expectation}, we see that the statistical uncertainty of the average sign is comparable to $S$ itself, and even negative values of $S_i$ are possible. Obviously, a (nearly) vanishing value of $S_i$ will lead to an either positive or negative spike in $V_i=(VS)_i/S_i$, which, in turn makes the resulting distribution $P(V_i)$ non-Gaussian.
In fact, it is well known~\cite{hatano} that the distribution of the ratio in Eq.~(\ref{eq:ratio}) is given by the superposition of a Gaussian and a Lorentzian, with the latter being responsible for the spikes.

From a practical perspective, this seems like bad news for multiple reasons: firstly, the outliers can be orders of magnitude away from the exact value, and, therefore, can be considered as meaningless; secondly, the variance as it is defined in Eq.~(\ref{eq:var}) only constitutes a meaningful measure for the statistical uncertainty for a normal distribution, but certainly not for the more complicated distribution observed in Fig.~\ref{fig:expectation2}. In other words, the associated Monte Carlo error bar becomes meaningless in this case.

\begin{figure}\centering
\includegraphics[width=0.485\textwidth]{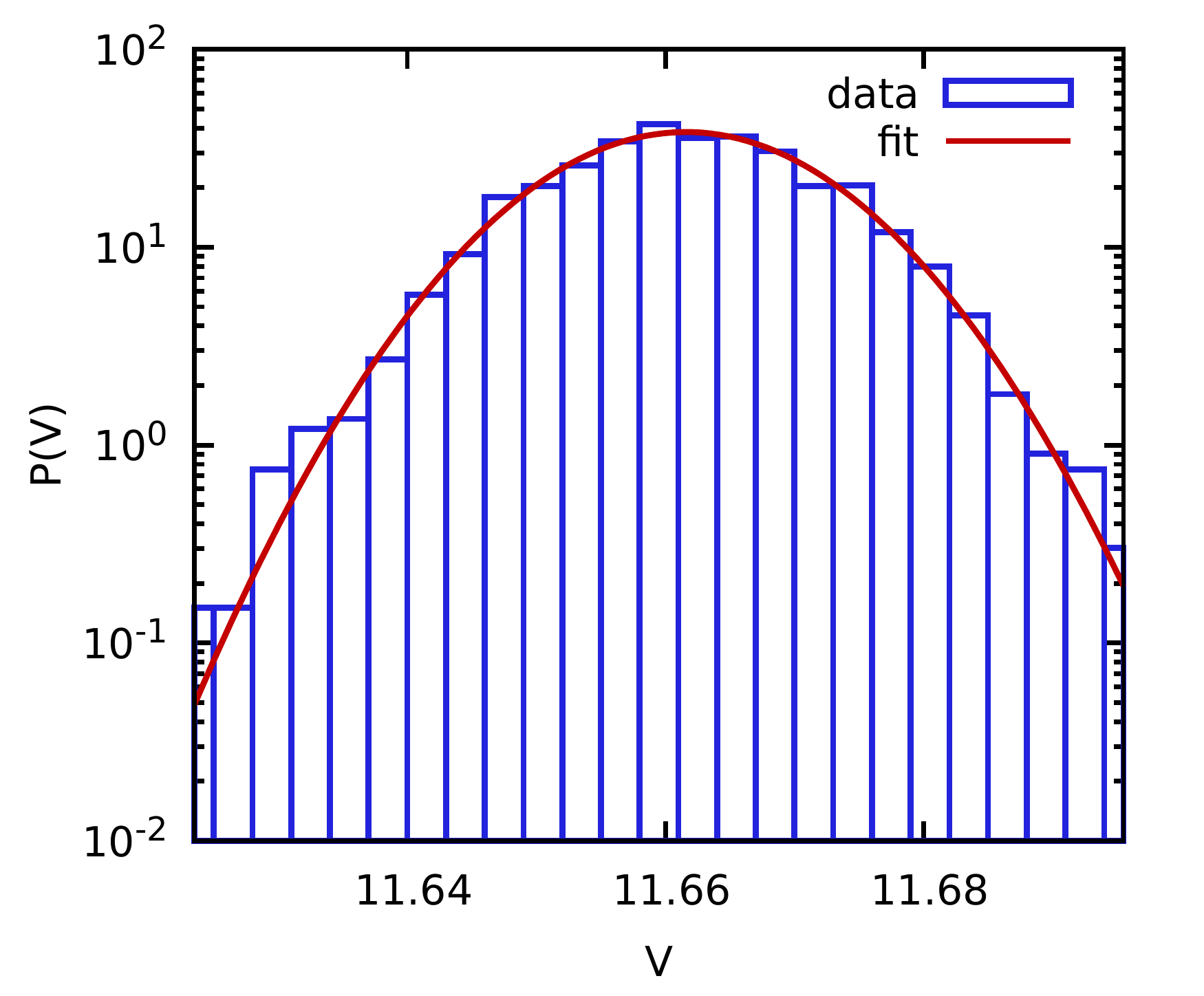}\includegraphics[width=0.485\textwidth]{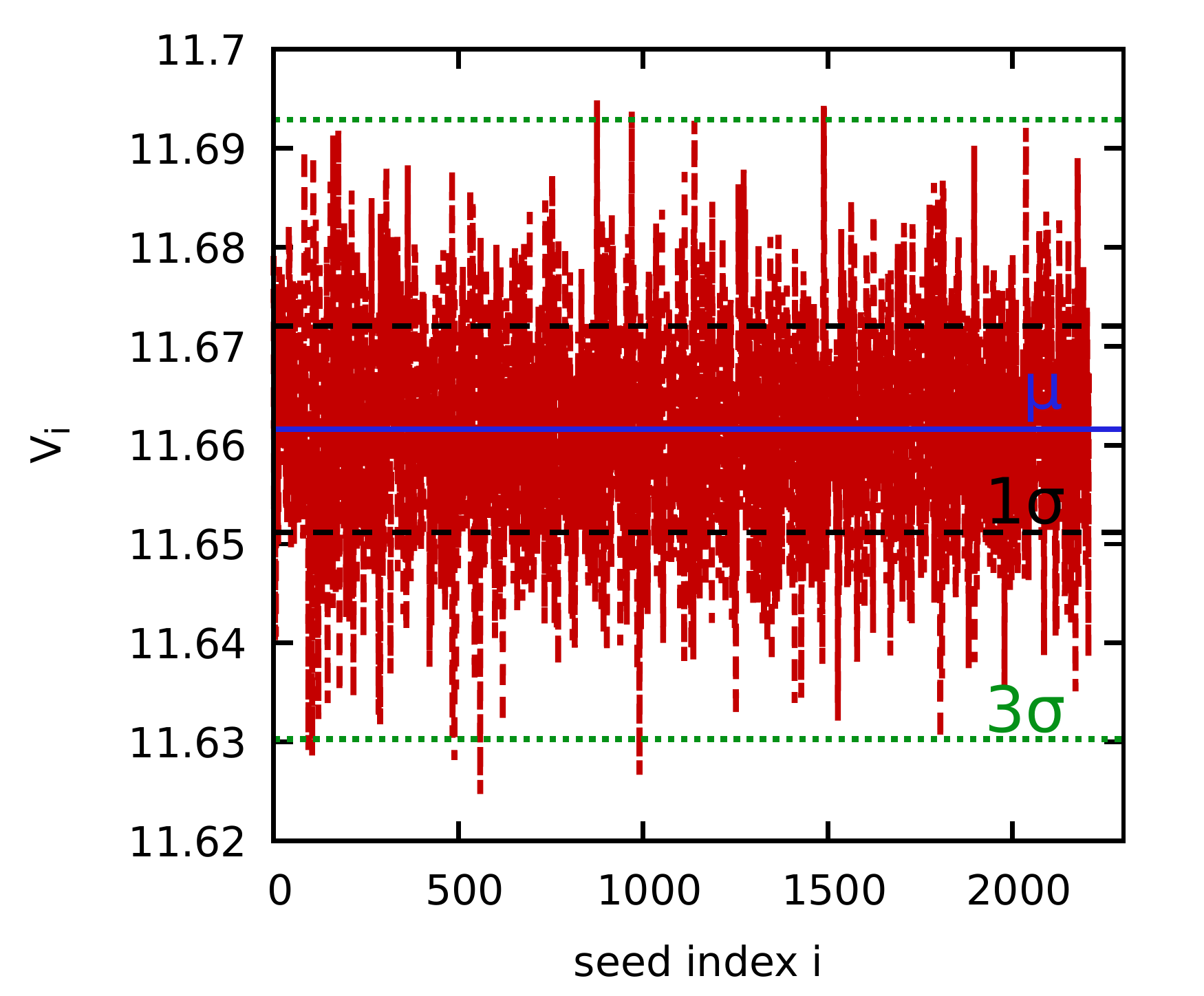}
\caption{\label{fig:expectation_lambda1}
Histograms of expectation values from $N\approx6000$ independent PIMC simulations with $N_\textnormal{MC}\approx4.8\times10^7$ measurements per simulation for a system with $\lambda=1$, $\mu=4.4$, and $\beta=1$. The left panel shows the distribution of the evaluation of the ratio from Eq.~(\ref{eq:ratio}) for each individual PIMC simulation, and the right panel the corresponding results for all of them, with the dashed black and dotted green lines indicating the $1\sigma$ and $3\sigma$ confidence intervals, and $\mu$ being the mean value.
}
\end{figure}

On the other hand, Dornheim~\cite{dornheim_sign_problem} has recently pointed out that this only becomes an issue when the relative error of the average sign in large. Conversely, if $S\gg\Delta S$, the skew in the distribution of Eq.~(\ref{eq:ratio}) becomes negligible and the distribution of the ratio should become indistinguishable from a Gaussian again. This is verified for the case of the GCE in the left panel of Fig.~\ref{fig:expectation_lambda1} for a larger value of the coupling strength, $\lambda=1$. For these parameters, we find $S\approx0.28$, and the statistical uncertainty $\Delta S$ of an individual PIMC simulation is approximately two orders of magnitude smaller. Consequently, the histogram of $V_i$ can indeed not be distinguished from a normal distribution and the fit is in excellent agreement to the data. Furthermore, the series of results for individual measurements shown in the right panel of the same figure appears to symmetrically fluctuate around the mean value without any spikes. More specifically, the blue horizontal line corresponds to the mean value $\mu$ from the Gaussian fit in the left panel, and the dashed black and dotted green lines to the $1\sigma$ and $3\sigma$ confidence intervals. There appear only $8$ individual $V_i$ outside of the latter, which would indeed be expected for $M\approx2300$ independent samples drawn from a Gaussian distribution.

In a nutshell, the estimation of an expectation value with a fermionic PIMC simulation is only quasi-exact when the relative statistical uncertainty in the average sign is small. This fact holds both in the CE and GCE.

\section{Summary and Discussion\label{sec:summary}}

In summary, we have presented a practical analysis of the fermion sign problem of fermionic PIMC simulations of electrons in $2D$ quantum dots in the GCE. Firstly, our implementation has been verified against independent CPIMC results by Schoof~\cite{Schoof} for different temperatures, and by checking the well-known virial theorem. 

Overall, we find that the sign problem is substantially more severe in the GCE compared to the CE at the same parameters. In particular, the grand-canonical sign is given by a superposition of the canonical signs for different $N$, and the latter decrease exponentially with the system size. Even worse, the canonical signs contribute not according to their importance for the fermionic expectation value of interest, but are weighted by their bosonic distribution $P(N)$. In other words, configurations with a large $N$ for which the fermionic distribution $P(N)$ is negligible thus still do substantially contribute to the sign problem in the GCE. In practice, this leads to a super-exponential decrease of the average sign (and, consequently, a super-exponential increase in the statistical uncertainty) both upon increasing the chemical potential $\mu$ and upon decreasing the coupling strength $\lambda$.

A more subtle problem of fermionic PIMC simulations in the GCE occurs at weak coupling, when the distributions of particle numbers $P(N)$ of bosons and fermions diverge. More specifically, the Pauli exclusion principle prohibits the multiple occupation of energetically low-lying states for Fermi-statistics, whereas there is no such restriction for bosons. Consequently, the bosonic distribution will be shifted to substantially larger values of $N$. This does not only lead to a drastic increase in the FSP compared to a canonical simulation with $N=\braket{\hat N}$ particles, but will also lead to a poor sampling of the relevant sectors themselves. In particular, the PIMC simulation will spend most time at configurations with large $N$ that do not contribute to fermionic observables at all, whereas configurations with small $N$ that would contribute to the latter might never be visited as they are not relevant to the bosonic reference system.

We thus conclude that fermionic PIMC simulations in the CE are in general more efficient compared to the GCE and, therefore, constitute the preferred option. Yet, grand-canonical PIMC simulations of fermions are still possible in some cases, and can give access to observables such as the compressibility or the Matsubara Greens function that cannot be obtained otherwise. This has interesting implications for many contemporary fields such as warm dense matter, ultracold atoms, or electrons in quantum dots. For example, the Matsubara Green function is connected to the single-particle spectral function $A(\mathbf{k},\omega)$ (with $\mathbf{k}$ and $\omega$ being the wave vector and frequency, respectively), which can be measured in experiments~\cite{nolting}.

\section*{Acknowledgments}
This work was partly funded by the Center of Advanced Systems Understanding (CASUS) which is financed by Germany's Federal Ministry of Education and Research (BMBF) and by the Saxon Ministry for Science, Culture and Tourism (SMWK) with tax funds on the basis of the budget approved by the Saxon State Parliament.
The PIMC calculations were carried out at the Norddeutscher Verbund f\"ur Hoch- und H\"ochstleistungsrechnen (HLRN) under grant shp00026, and on a Bull Cluster at the Center for Information Services and High Performace Computing (ZIH) at Technische Universit\"at Dresden.

\section*{References}

\bibliographystyle{iopart-num}
\bibliography{bibliography.bib}

\providecommand{\newblock}{}
\begin{thebibliography}{10}
\expandafter\ifx\csname url\endcsname\relax
  \def\url#1{{\tt #1}}\fi
\expandafter\ifx\csname urlprefix\endcsname\relax\def\urlprefix{URL }\fi
\providecommand{\eprint}[2][]{\url{#2}}

\bibitem{Fosdick_PR_1966}
Fosdick L~D and Jordan H~F 1966 {\em Phys. Rev.\/} {\bf 143}(1) 58--66
  \urlprefix\url{https://link.aps.org/doi/10.1103/PhysRev.143.58}

\bibitem{Jordan_PR_1968}
Jordan H~F and Fosdick L~D 1968 {\em Phys. Rev.\/} {\bf 171}(1) 128--149
  \urlprefix\url{https://link.aps.org/doi/10.1103/PhysRev.171.128}

\bibitem{Berne_JCP_1982}
Herman M~F, Bruskin E~J and Berne B~J 1982 {\em The Journal of Chemical
  Physics\/} {\bf 76} 5150--5155 (\textit{Preprint}
  \eprint{https://doi.org/10.1063/1.442815})
  \urlprefix\url{https://doi.org/10.1063/1.442815}

\bibitem{Takahashi_Imada_PIMC_1984}
Takahashi M and Imada M 1984 {\em Journal of the Physical Society of Japan\/}
  {\bf 53} 963--974

\bibitem{Pollock_Ceperley_PRB_1984}
Pollock E~L and Ceperley D~M 1984 {\em Phys. Rev. B\/} {\bf 30}(5) 2555--2568
  \urlprefix\url{https://link.aps.org/doi/10.1103/PhysRevB.30.2555}

\bibitem{cep}
Ceperley D~M 1995 {\em Rev. Mod. Phys\/} {\bf 67} 279
  \urlprefix\url{https://journals.aps.org/rmp/abstract/10.1103/RevModPhys.67.279}

\bibitem{ultracold2}
Pollock E~L and Ceperley D~M 1987 {\em Phys. Rev. Lett\/} {\bf 36} 8343
  \urlprefix\url{https://journals.aps.org/prb/abstract/10.1103/PhysRevB.36.8343}

\bibitem{Dornheim_PRB_2015}
Dornheim T, Filinov A and Bonitz M 2015 {\em Phys. Rev. B\/} {\bf 91}(5) 054503
  \urlprefix\url{https://link.aps.org/doi/10.1103/PhysRevB.91.054503}

\bibitem{Kwon_PRL_2002}
Kwon Y and Whaley K~B 2002 {\em Phys. Rev. Lett.\/} {\bf 89}(27) 273401
  \urlprefix\url{https://link.aps.org/doi/10.1103/PhysRevLett.89.273401}

\bibitem{Saito}
Saito H 2016 {\em Journal of the Physical Society of Japan\/} {\bf 85} 053001
  (\textit{Preprint} \eprint{https://doi.org/10.7566/JPSJ.85.053001})
  \urlprefix\url{https://doi.org/10.7566/JPSJ.85.053001}

\bibitem{PhysRevLett.56.351}
Ceperley D~M and Pollock E~L 1986 {\em Phys. Rev. Lett.\/} {\bf 56}(4) 351--354
  \urlprefix\url{https://link.aps.org/doi/10.1103/PhysRevLett.56.351}

\bibitem{Saccani_Supersolid_PRL_2012}
Saccani S, Moroni S and Boninsegni M 2012 {\em Phys. Rev. Lett.\/} {\bf
  108}(17) 175301
  \urlprefix\url{https://link.aps.org/doi/10.1103/PhysRevLett.108.175301}

\bibitem{Filinov_PRA_2012}
Filinov A and Bonitz M 2012 {\em Phys. Rev. A\/} {\bf 86}(4) 043628
  \urlprefix\url{https://link.aps.org/doi/10.1103/PhysRevA.86.043628}

\bibitem{dornheim_dynamic}
Dornheim T, Groth S, Vorberger J and Bonitz M 2018 {\em Phys. Rev. Lett.\/}
  {\bf 121} 255001
  \urlprefix\url{https://journals.aps.org/prl/abstract/10.1103/PhysRevLett.121.255001}

\bibitem{PhysRevB.98.134509}
Kora Y and Boninsegni M 2018 {\em Phys. Rev. B\/} {\bf 98}(13) 134509
  \urlprefix\url{https://link.aps.org/doi/10.1103/PhysRevB.98.134509}

\bibitem{review}
Dornheim T, Groth S and Bonitz M 2018 {\em Phys. Reports\/} {\bf 744} 1--86
  \urlprefix\url{https://www.sciencedirect.com/science/article/abs/pii/S0370157318300516}

\bibitem{new_POP}
Bonitz M, Dornheim T, Moldabekov Z~A, Zhang S, Hamann P, Kählert H, Filinov A,
  Ramakrishna K and Vorberger J 2020 {\em Physics of Plasmas\/} {\bf 27} 042710
  (\textit{Preprint} \eprint{https://doi.org/10.1063/1.5143225})
  \urlprefix\url{https://doi.org/10.1063/1.5143225}

\bibitem{wdm_book}
Graziani F, Desjarlais M~P, Redmer R and Trickey S~B (eds) 2014 {\em Frontiers
  and Challenges in Warm Dense Matter\/} (International Publishing: Springer)

\bibitem{fortov_review}
Fortov V~E 2009 {\em Phys.-Usp\/} {\bf 52} 615--647

\bibitem{Egger_PRL_1999}
Egger R, H\"ausler W, Mak C~H and Grabert H 1999 {\em Phys. Rev. Lett.\/} {\bf
  82}(16) 3320--3323
  \urlprefix\url{https://link.aps.org/doi/10.1103/PhysRevLett.82.3320}

\bibitem{RevModPhys.74.1283}
Reimann S~M and Manninen M 2002 {\em Rev. Mod. Phys.\/} {\bf 74}(4) 1283--1342
  \urlprefix\url{https://link.aps.org/doi/10.1103/RevModPhys.74.1283}

\bibitem{Dornheim_PRA_2020}
Dornheim T 2020 {\em Phys. Rev. A\/} {\bf 102}(2) 023307
  \urlprefix\url{https://link.aps.org/doi/10.1103/PhysRevA.102.023307}

\bibitem{boninsegni1}
Boninsegni M, Prokofev N~V and Svistunov B~V 2006 {\em Phys. Rev. E\/} {\bf 74}
  036701
  \urlprefix\url{https://journals.aps.org/pre/abstract/10.1103/PhysRevE.74.036701}

\bibitem{boninsegni2}
Boninsegni M, Prokofev N~V and Svistunov B~V 2006 {\em Phys. Rev. Lett\/} {\bf
  96} 070601
  \urlprefix\url{https://journals.aps.org/prl/abstract/10.1103/PhysRevLett.96.070601}

\bibitem{Loh_sign_problem_PRB_1990}
Loh E~Y, Gubernatis J~E, Scalettar R~T, White S~R, Scalapino D~J and Sugar R~L
  1990 {\em Phys. Rev. B\/} {\bf 41}(13) 9301--9307
  \urlprefix\url{https://link.aps.org/doi/10.1103/PhysRevB.41.9301}

\bibitem{dornheim_sign_problem}
Dornheim T 2019 {\em Phys. Rev. E\/} {\bf 100} 023307
  \urlprefix\url{https://journals.aps.org/pre/abstract/10.1103/PhysRevE.100.023307}

\bibitem{troyer}
Troyer M and Wiese U~J 2005 {\em Phys. Rev. Lett\/} {\bf 94} 170201
  \urlprefix\url{http://link.aps.org/doi/10.1103/PhysRevLett.94.170201}

\bibitem{Brown_PRL_2013}
Brown E~W, Clark B~K, DuBois J~L and Ceperley D~M 2013 {\em Phys. Rev. Lett.\/}
  {\bf 110}(14) 146405
  \urlprefix\url{https://link.aps.org/doi/10.1103/PhysRevLett.110.146405}

\bibitem{Blunt_PRB_2014}
Blunt N~S, Rogers T~W, Spencer J~S and Foulkes W~M~C 2014 {\em Phys. Rev. B\/}
  {\bf 89}(24) 245124
  \urlprefix\url{https://link.aps.org/doi/10.1103/PhysRevB.89.245124}

\bibitem{Dornheim_NJP_2015}
Dornheim T, Groth S, Filinov A and Bonitz M 2015 {\em New Journal of Physics\/}
  {\bf 17} 073017
  \urlprefix\url{https://doi.org/10.1088\%2F1367-2630\%2F17\%2F7\%2F073017}

\bibitem{Blunt_PRL_2015}
Blunt N~S, Alavi A and Booth G~H 2015 {\em Phys. Rev. Lett.\/} {\bf 115}(5)
  050603
  \urlprefix\url{https://link.aps.org/doi/10.1103/PhysRevLett.115.050603}

\bibitem{Malone_PRL_2016}
Malone F~D, Blunt N~S, Brown E~W, Lee D~K~K, Spencer J~S, Foulkes W~M~C and
  Shepherd J~J 2016 {\em Phys. Rev. Lett.\/} {\bf 117}(11) 115701
  \urlprefix\url{https://link.aps.org/doi/10.1103/PhysRevLett.117.115701}

\bibitem{dornheim_prl}
Dornheim T, Groth S, Sjostrom T, Malone F~D, Foulkes W~M~C and Bonitz M 2016
  {\em Phys. Rev. Lett.\/} {\bf 117} 156403
  \urlprefix\url{http://link.aps.org/doi/10.1103/PhysRevLett.117.156403}

\bibitem{groth_prl}
Groth S, Dornheim T, Sjostrom T, Malone F~D, Foulkes W~M~C and Bonitz M 2017
  {\em Phys. Rev. Lett.\/} {\bf 119} 135001
  \urlprefix\url{https://journals.aps.org/prl/abstract/10.1103/PhysRevLett.119.135001}

\bibitem{Yilmaz_JCP_2020}
Yilmaz A, Hunger K, Dornheim T, Groth S and Bonitz M 2020 {\em The Journal of
  Chemical Physics\/} {\bf 153} 124114 (\textit{Preprint}
  \eprint{https://doi.org/10.1063/5.0022800})
  \urlprefix\url{https://doi.org/10.1063/5.0022800}

\bibitem{Lee_JCP_2021}
Lee J, Morales M~A and Malone F~D 2021 {\em The Journal of Chemical Physics\/}
  {\bf 154} 064109 (\textit{Preprint}
  \eprint{https://doi.org/10.1063/5.0041378})
  \urlprefix\url{https://doi.org/10.1063/5.0041378}

\bibitem{Dornheim_PRL_2020}
Dornheim T, Vorberger J and Bonitz M 2020 {\em Phys. Rev. Lett.\/} {\bf 125}(8)
  085001
  \urlprefix\url{https://link.aps.org/doi/10.1103/PhysRevLett.125.085001}

\bibitem{dornheim_ML}
Dornheim T, Vorberger J, Groth S, Hoffmann N, Moldabekov Z and Bonitz M 2019
  {\em J. Chem. Phys\/} {\bf 151} 194104
  \urlprefix\url{https://aip.scitation.org/doi/full/10.1063/1.5123013}

\bibitem{Rubenstein_auxiliary_finite_T}
Liu Y, Cho M and Rubenstein B 2018 {\em Journal of Chemical Theory and
  Computation\/} {\bf 14} 4722--4732 ISSN 1549-9618
  \urlprefix\url{https://doi.org/10.1021/acs.jctc.8b00569}

\bibitem{metropolis}
Metropolis N, Rosenbluth A~W, Rosenbluth M~N, Teller A~H and Teller E 1953 {\em
  The Journal of Chemical Physics\/} {\bf 21} 1087--1092 (\textit{Preprint}
  \eprint{https://doi.org/10.1063/1.1699114})
  \urlprefix\url{https://doi.org/10.1063/1.1699114}

\bibitem{Dornheim_CPP_2019}
Dornheim T, Groth S and Bonitz M 2019 {\em Contributions to Plasma Physics\/}
  {\bf 59} e201800157 (\textit{Preprint}
  \eprint{https://onlinelibrary.wiley.com/doi/pdf/10.1002/ctpp.201800157})
  \urlprefix\url{https://onlinelibrary.wiley.com/doi/abs/10.1002/ctpp.201800157}

\bibitem{dornheim_POP}
Dornheim T, Groth S, Malone F~D, Schoof T, Sjostrom T, Foulkes W~M~C and Bonitz
  M 2017 {\em Physics of Plasmas\/} {\bf 24} 056303 (\textit{Preprint}
  \eprint{https://doi.org/10.1063/1.4977920})
  \urlprefix\url{https://doi.org/10.1063/1.4977920}

\bibitem{dornheim_permutation_cycles}
Dornheim T, Groth S, Filinov A~V and Bonitz M 2019 {\em The Journal of Chemical
  Physics\/} {\bf 151} 014108 \urlprefix\url{https://doi.org/10.1063/1.5093171}

\bibitem{krauth2006statistical}
Krauth W 2006 {\em Statistical Mechanics: Algorithms and Computations\/} Oxford
  Master Series in Physics (Oxford University Press, UK) ISBN 9780198515357
  \urlprefix\url{https://books.google.de/books?id=B3koVucDyKUC}

\bibitem{https://doi.org/10.1002/ctpp.201100012}
Schoof T, Bonitz M, Filinov A, Hochstuhl D and Dufty J 2011 {\em Contributions
  to Plasma Physics\/} {\bf 51} 687--697 (\textit{Preprint}
  \eprint{https://onlinelibrary.wiley.com/doi/pdf/10.1002/ctpp.201100012})
  \urlprefix\url{https://onlinelibrary.wiley.com/doi/abs/10.1002/ctpp.201100012}

\bibitem{doi:10.1063/5.0030760}
Dornheim T, Invernizzi M, Vorberger J and Hirshberg B 2020 {\em The Journal of
  Chemical Physics\/} {\bf 153} 234104 (\textit{Preprint}
  \eprint{https://doi.org/10.1063/5.0030760})
  \urlprefix\url{https://doi.org/10.1063/5.0030760}

\bibitem{doi:10.1063/5.0008720}
Hirshberg B, Invernizzi M and Parrinello M 2020 {\em The Journal of Chemical
  Physics\/} {\bf 152} 171102 (\textit{Preprint}
  \eprint{https://doi.org/10.1063/5.0008720})
  \urlprefix\url{https://doi.org/10.1063/5.0008720}

\bibitem{Schoof}
Schoof T 2016 {\em Configuration Path Integral Monte Carlo: Ab initio
  simulations of fermions in the warm dense matter regime\/} Ph.D. thesis Kiel
  university

\bibitem{Groth_PRB_2016}
Groth S, Schoof T, Dornheim T and Bonitz M 2016 {\em Phys. Rev. B\/} {\bf
  93}(8) 085102
  \urlprefix\url{https://link.aps.org/doi/10.1103/PhysRevB.93.085102}

\bibitem{Schoof_CPP_2015}
Schoof T, Groth S and Bonitz M 2015 {\em Contributions to Plasma Physics\/}
  {\bf 55} 136--143 (\textit{Preprint}
  \eprint{https://onlinelibrary.wiley.com/doi/pdf/10.1002/ctpp.201400072})
  \urlprefix\url{https://onlinelibrary.wiley.com/doi/abs/10.1002/ctpp.201400072}

\bibitem{Ott2018}
Ott T, Thomsen H, Abraham J~W, Dornheim T and Bonitz M 2018 {\em The European
  Physical Journal D\/} {\bf 72} 84 ISSN 1434-6079
  \urlprefix\url{https://doi.org/10.1140/epjd/e2018-80385-7}

\bibitem{greiner1995thermodynamics}
Greiner W, Neise L, Stocker H, St{\"o}cker H and Rischke D 1995 {\em
  Thermodynamics and Statistical Mechanics\/} Classical theoretical physics
  (Springer-Verlag) ISBN 9780387942995
  \urlprefix\url{https://books.google.de/books?id=V6zvAAAAMAAJ}

\bibitem{Janke_JCP_1997}
Janke W and Sauer T 1997 {\em The Journal of Chemical Physics\/} {\bf 107}
  5821--5839 (\textit{Preprint} \eprint{https://doi.org/10.1063/1.474309})
  \urlprefix\url{https://doi.org/10.1063/1.474309}

\bibitem{landau}
Zagrebnov V~A and Bru J~B 2001 {\em Physics Reports\/} {\bf 350} 291--434 ISSN
  0370-1573
  \urlprefix\url{https://www.sciencedirect.com/science/article/pii/S0370157300001320}

\bibitem{hatano}
Hatano N 1994 {\em Journal of the Physical Society of Japan\/} {\bf 63}
  1691--1697 (\textit{Preprint} \eprint{https://doi.org/10.1143/JPSJ.63.1691})
  \urlprefix\url{https://doi.org/10.1143/JPSJ.63.1691}

\bibitem{nolting}
Nolting W and Brewer W~D 2009 {\em Fundamentals of Many-body Physics:
  Principles and Methods\/} (Heidelberg: Springer)

\end{thebibliography}

\end{document}